\documentclass[preprintnumbers,amsmath,amssymb,floatfix,10pt,prd,onecolumn,
superscriptaddress,nofootinbib]{revtex4}

\usepackage{amsfonts}
\usepackage{mathrsfs}
\usepackage{latexsym}
\usepackage{epsfig}
\usepackage{epstopdf}
\usepackage{graphicx}
\usepackage{amssymb}
\usepackage{amsmath}
\usepackage{dcolumn}
\usepackage{bm}
\usepackage{color}
\usepackage{xcolor}
\usepackage{comment}
\usepackage[cp1251]{inputenc}
\begin{document}

\title{\bf Impact of Ricci Inverse Gravity on Hybrid Star Model}

\author{Adnan Malik}
\email{adnan.malik@zjnu.edu.cn; adnan.malik@skt.umt.edu.pk; adnanmalik_chheena@yahoo.com}
\affiliation{School of Mathematical Sciences, Zhejiang Normal University, \\Jinhua, Zhejiang, China.}
\affiliation{Department of Mathematics, University of Management and Technology,\\ Sialkot Campus, Lahore, Pakistan}

\author{Amjad Hussain}
\email{amjad.haider@nu.edu.pk}
\affiliation{National University of Computer and Emerging Sciences,\\Chiniot-Faisalabad Campus, Pakistan}

\author{M. Farasat Shamir}
\email{farasat.shamir@nu.edu.pk}
\affiliation{National University of Computer and Emerging Sciences,\\ Lahore Campus, Pakistan.}

\author{Ayesha Almas}
\email{ayesha787@icloud.com}\affiliation{Department of Mathematics, University of Management and Technology,\\ Sialkot Campus, Pakistan.}

\begin{abstract}
The objective of our current study is to explore novel aspects of a stationary anisotropic relativistic hybrid compact star that consists of quark matter (QM) in its core and ordinary baryonic matter (OBM) in its crust. This study has been done by adopting separate equations of states (EoSs) for quark matter and baryonic matter. The MIT bag model equation of state $p_{q}=\frac{1}{3}(\rho_{q}-4B)$ has been used to demonstrate a correlation between the density and pressure of weird quark matter in the interior of the star. In addition, we present a simple linear equation of state $p_{r}=\beta_{1}\rho-\beta$ that links radial pressure and matter density for OBM. The stellar model was formulated within the context of $f(\mathcal{R},\mathcal{A})$ gravity, utilizing a linear correlation between Ricci tensor $\mathcal{R}$ and anticurvature scaler $\mathcal{A}$.
To solve the field equations of this novel alternative gravity, we employ the Krori and Barua approach to the metric potentials. The validity of our suggested model is assessed using the  graphical approach, while ensuring that the conditions are physically feasible. Our focus is specifically on the tiny celestial object known as LMC X-4 [$\text{M} = (1.04^{+0.09}_{-0.09})M_{\odot};\text{R}=8.301^{+0.2}_{-0.2}\text{km}$], which we consider a viable candidate for a strange quark star. We want to clarify the model's physical validity by examining a variety of physical assessments, including dynamical equilibrium, energy conditions, compactness factor, mass function, and surface redshift. The resulting outcome confirms the authenticity of the hybrid star model under analysis.
\end{abstract}

\maketitle

\date{\today}
\section{INTRODUCTION}
	
Once a star exhausts its fuel, it usually undergoes a series of changes that result in the creation of a dense object such as a white dwarf, neutron star, or black hole, which is determined by its mass. Neutron stars are famous for their exceptional compactness and density. The compact stars have a mass ranging from $1\sim2$ $M_{\odot}$ and a very small radius of $10\sim13$ km. As a result, the average density of matter within these dense stars is around $10^{14}$ $g/cm^3$.
Through the use of electromagnetic spectrum and gravitational wave observations, we have gathered a wealth of information about these dense celestial objects in recent years. Neutron stars are created when huge stars have supernova explosions and gravitational collapse, causing their cores to compress to densities similar to atomic nuclei. They themselves can take various forms due to the high density in the core; for instance they could be composed of normal matter mixed with hyperons and/or condensed mesons. At such high densities, the matter inside the star has the potential to convert into quark matter. Numerous studies have been conducted to support the hypothesis that massive neutron stars could contain quark matter \cite{PR8,PR9,PR10,PR11,PR12,PR13}. Observations have shown that the existence of free quark matter could potentially give rise to a new class of compact stars that have a core made up of QM surrounded by an outer layer made up of OBM \cite{Implications-of-supermassive-neutron-star36, Implications-of-supermassive-neutron-star37, Implications-of-supermassive-neutron-star38}. These stars are named hybrid stars, and they are different from conventional neutron stars, which consist solely of hadronic matter \cite{Implications-of-supermassive-neutron-star39, Implications-of-supermassive-neutron-star40, Implications-of-supermassive-neutron-star41, Implications-of-supermassive-neutron-star42, Implications-of-supermassive-neutron-star43}. Researchers have investigated the possible existence of these hybrid stars, which have masses similar to the mass of the second object in GW190814, and explored the properties of such massive hybrid stars \cite{Implications-of-supermassive-neutron-star44, Implications-of-supermassive-neutron-star45, Implications-of-supermassive-neutron-star46, Implications-of-supermassive-neutron-star47, Implications-of-supermassive-neutron-star48}.

The inquiry about the rapid expansion of the cosmos in General Relativity (GR) can be addressed by modifying the lagrangian of GR, leading to the formulation of modified theories of gravity (MTG). These MTG extend beyond GR as they introduce additional degrees of freedom and higher-order derivatives in the field equations. Although GR has achieved success in demonstrating gravity on various scales \cite{S1, S2}, the need for modification arises due to its limited capacity to comprehensively account for phenomena such as dark energy and the universe' accelerating expansion. So in the frame work of MTG, researchers aim to bridge the gap left by GR for comprehending the dynamics of the cosmos. Reference \cite{S10} provides a historical review of the early efforts to broaden the scope of GR, driven by the various motivations. For a thorough understanding of MTG, see references \cite{S11,S12}. This study focuses on Ricci-inverse gravity \cite{S13} as an alternative theory of gravity in relation to the investigation of hybrid stars. A unique class of fourth-order MTG model is introduced in this framework by the generalization of the Einstein-Hilbert action. The novel alternative gravity proposed by Amendola et al. \cite{S13} involves the generalization of Einstein-Hilbert action. The reference \cite{S14} provides a generalization of Ricci inverse gravity and establishes two distinct classes of Ricci inverse gravity: Class I and Class II. In Class I, the Lagrangian incorporates an arbitrary function $f$ that is dependent on the Ricci scaler $\mathcal{R}$ and the anticurvature scalar $\mathcal{A}$, defined as the trace of the Ricci-inverse tensor $A_{\mu\nu}$. It is important to remember that the anti-curvature scalar $\mathcal{A}$ does not correspond to the inverse of the Ricci scalar $\mathcal{R}$. On the other hand, in Class II, the function takes the form $f (\mathcal{R}, A^{\mu\nu} A_{\mu\nu})$, constituting a function of the Ricci scalar and the square of the anti-curvature tensor. Researchers have investigated Ricci-inverse gravity in several contexts. The utilization of Ricci inverse gravity to investigate the compact star structure in the context of spherically symmetric spacetime was initially introduced by Shamir et al. \cite{M40}. An analysis has been conducted on the matter-antimatter asymmetry through baryogenesis in the realm of $f (\mathcal{R}, \mathcal{A})$ \cite{S16}. Additionally, a study has been carried out on a non-relativistic static and spherically symmetric cosmic structure embedded into a de Sitter cosmology\cite{S17}. Furthermore, researchers have investigated a no-go theorem for inflation within an extended Ricci-inverse gravity model \cite{S18,S19}, emphasizing the various uses and consequences of MTG in cosmological and astrophysical scenarios. The study by Souza et al. \cite{M42} calculated two different axially symmetric spacetimes, and they showed that Ricci inverse gravity allows the existence of closed time like curves (CTCs).

There is no astrophysical entity that is entirely composed of perfect fluid. Therefore, the incorporation of pressure anisotropy is crucial in the formulation of accurate models for ultra dense stars, since it significantly impacts their structure and characteristics. The incorporation of anisotropy has a substantial impact on observable properties, including the ratio of mass to radius, moment of inertia, redshift, and stability of star configurations. Ruderman \cite{PR56} proposed through theoretical studies that stellar matter can display anisotropy when its density exceeds $10^{15}$ $gm/cm^3$. For such high-dense stellar structures, there can be a disparity between radial pressure $p_{r}$ and tangential pressure $p_{t}$, and both components are functions of radial coordinate r due to the spherical symmetry. The primary causes that contribute to pressure anisotropy in the distribution of matter are viscosity, pion condensation \cite{PR62}, phase transitions in superfluids \cite{PR63}, or other related physical processes. These factors can induce anisotropy in the pressure within super dense stars, impacting their internal structure and properties. Mak and Harko \cite{PR65} conducted computational simulations to model celestial bodies with anisotropy using the principles of general relativity. In this context, \cite{PR68,PR69,PR70,PR71,PR72} encompass further relevant studies concerning anisotropic compact stellar objects.

Building on the observations we've already made, our goal is to do a full investigation that gives a complete picture of how the physical features of relativistic hybrid neutron stars change when anisotropy is present. Our current work is expected to provide significant findings for the examination of compact stellar entities, which will also enabling us to study the exploration of matter's response under intense gravitational forces. The thermodynamic bag model (tdBag) and the Nambu-Jona-Lasino type (NJL) models are the two most often used effective quark matter models in astrophysics \cite{PR74,PR75,PR76,PR77}. The thermodynamic bag model (tdBAG) is widely used in the field of astrophysics compared to the other models. The recently introduced vBag model \cite{PR78} have been shown to be a useful model for conducting astrophysical investigations. In this paper, we use the MIT bag model as an EoS for QM. Numerous researchers have effectively utilized this MIT bag EoS model to simulate strange quark stars (SQS) \cite{PR79,PR80,PR81,PR82,PR83,PR84,PR85}. This phenomenological model is the initial successful representation of the phase transition from the quark gluon plasma (QGP) state to the hadronic phase. The pressure-density relation for QM in the MIT bag EoS is expressed as $p_{q}=\frac{1}{3}(\rho_{q}-4B_{g})$, where $\rho_q$ and $p_q$ represent the density and pressure of quark matter, respectively. The bag constant $B_g$ in this context is defined as the difference between the densities of perturbative and non-perturbative quantum chromo dynamical (QCD) vacuums. The fundamental model is well-suited for studying the stable arrangement of a cosmic compact object made up of u, d, and s quarks. Usually, the $B_{g}$ falls within the range of 58.9-91.5 $MeV/fm^3$ for massless strange quarks, while for massive quarks $B_{g}$ ranges of 56-78 $MeV/fm^3$ \cite{PR86}.

Researchers have shown significant attention to investigating the stellar structures by employing the Krori-Barua metric (KB metric) measure. Bhar \cite{F35} uses KB metric potentials to study a unique hybrid star model that incorporates OBM and QM in the interior region of stars. In their study, Rahaman et al. \cite{F36} investigated the feasibility of utilizing the KB model for the purpose of characterizing ultra-compact objects, such as strange stars. They also conducted an analysis of the implications of employing a mathematical representation for the purpose of modeling these strange stellar phenomena. Biswas et al. \cite{F37} explored the anisotropic charged strange stars by utilizing the KB spacetime in the framework of the $f (\mathcal{R}, \mathcal{T})$ theory of gravity. Sharif and Waseem \cite{F38} studied the behavior of anisotropic compact stars under the $\mathcal{R} + \alpha R_{\mu\nu} \mathcal{T}^{ \mu\nu}$ gravity theory model. The KB metric was used for the investigation and determined the parameters's value by utilizing the observed masses and radii of compact stars such as Her X-1, SAX J 1808.43658, and 4U182030. Using the Gauss-Bonnet MTG, also referred to as $f (\mathcal{G})$ MTG, Abbas et al. \cite{F39} explored the development of compact stars in the presence of anisotropy. This theory is widely regarded as a leading candidate for explaining the rapid expansion of the cosmos. They utilized an analytical solution obtained from the KB metric, which was applied to the Einstein field equations. Their analysis utilized an anisotropic form of matter and a power law representation of $f (\mathcal{G})$ gravity. Considering the utilization of KB metric potentials , Shamir and Malik \cite{F40} examined the characteristics of anisotropic compact stars within the framework of the $f (\mathcal{R}, \phi)$ theory of gravity. In this context, $\mathcal{R}$ represents the Ricci scalar, while $\phi$ represents the scalar field. Malik et al. \cite{F41,F42}, investigated the behavior of compact stars in the presence of charge in the modified $f (\mathcal{R})$ theories of gravity. They utilized the KB metric potentials and presented a singularity-free solution for a charged fluid sphere. Incorporating the Chaplygin EoS in $f(\mathcal{R},\mathcal{T})$ MTG, Bhar \cite{F43} studied the spherically symmetric spacetime along with anisotropic fluid distribution in the presence of an electric field. Hossein et al. \cite{F44} employed the KB metric potentials to model the formation of anisotropic compact star. They achieved this by considering the analytical solution of the metric potentials. Subsequently, Rehman et al. \cite{F45} investigated the feasibility of utilizing the Krori and Barua model to characterize ultra-compact objects such as strange stars. They also established some limitations on the model parameters to ensure its stability when studying strange stars.\\

Our paper is structured as follows: Section II introduces the fundamental concepts of Ricci inverse gravity theory, along with the model used for $f(\mathcal{R},\mathcal{A})$. Section III provides a detailed explanation of the energy-momentum tensor components, field equations, and EoSs for both OBM and QM. Section IV details the analytic expressions of $\rho$, $p_{r}$, $p_{t}$, $\rho_{q}$, and $p_{q}$ using KB metric potentials. This section also discusses the methodology  for determining the unknown parameters present in the metric potentials by lining up the interior spacetime with the exterior Schwarzschild line element. Section V delves into the graphical analysis of various physical characteristics of the hybrid star model, such as energy density, radial and tangential pressure components, anisotropy, energy conditions, and EOS parameters, mass function, compactness factor, and redshift function. Section VI focuses on assessing our model's stability by examining the speed of sound, adiabatic index, and equilibrium condition of all forces. Finally, section VII summarizes our findings.

\section{Basic formalism of Ricci Inverse Gravity}
This section presents a concise summary of the basic principles of Ricci inverse gravity, a newly formulated alternative theory of gravity. The inverse of the Ricci tensor $\mathcal{R}_{\mu\nu}$ is referred to as the anticurvature tensor $\mathcal{A}^{\mu\nu}$. The relationship between the anticurvature tensor and the Ricci tensor can be expressed as
\begin{equation}\label{a1}
	\mathcal{A}^{\mu\nu}\mathcal{R}_{\nu\sigma}=\delta^{\mu}_{\sigma} ,
\end{equation}
where $\delta^{\mu}_{\sigma}$ is called Kronecker delta. The anti-curvature scalar $\mathcal{A}$ which is based on the anti-curvature tensor can be defined as
\begin{equation}\label{a2}	
	\mathcal{A}=g_{\mu\nu}\mathcal{A}^{\mu\nu},
\end{equation}
(obviously, $\mathcal{A}\neq\mathcal{R}^{-1}$).
It should be noted that in the case, where the determinant of Ricci tensor is zero, there is no way to invert $\mathcal{R_{\mu\nu}}$. Hence the inclusion of $\text{det}(\mathcal{R}_{\mu\nu})\neq 0$ is an essential prerequisite to investigate a metric as a potential solution of Ricci-inverse gravity.

To formulate the field equations of modified Ricci inverse gravity, one can start by considering the following action:
\begin{equation}\label{a3}
	S=\int d^4x\sqrt{-g}f(\mathcal{R},\mathcal{A})+\int d^4x\sqrt{-g}\mathcal{L}_{m},
\end{equation}
here, $f(\mathcal{R},\mathcal{A})$ represents a function that can take any values for $\mathcal{R}$ and $\mathcal{A}$, $\mathcal{L}_{m}$ refers to the matter langrangian, while $g$ represents the determinant of the metric tensor. The field equations are obtained by calculating the variation of the action defined in Eq. (\ref{a3}) with respect to the metric tensor, resulting in the following set of equations \cite{Amendola}.

\begin{equation}\label{a4}
	 f_{\mathcal{R}}\mathcal{R}^{\mu\nu}-f_{\mathcal{A}}\mathcal{A}^{\mu\nu}-\frac{1}{2}fg^{\mu\nu}+g^{\gamma\mu}\nabla_{\delta}\nabla_{\gamma}f_{\mathcal{A}}\mathcal{A}^{\delta}_{\zeta}\mathcal{A}^{\nu\zeta}-\frac{1}{2}\nabla^{2}(f_{\mathcal{A}}\mathcal{A}^{\mu}_{\zeta}\mathcal{A}^{\nu\zeta})-\frac{1}{2}g^{\mu\nu}\nabla_{\delta}\nabla_{\eta}(f_{\mathcal{A}}\mathcal{A}^{\delta}_{\zeta}\mathcal{A}^{\eta\zeta})-\nabla^{\mu}\nabla^{\nu}f_{\mathcal{R}}+g^{\mu\nu}\nabla^{2}f_{\mathcal{R}}=\kappa\mathcal{T}^{\mu\nu},
\end{equation}
where $f_{\mathcal{R}}=\frac{\partial f}{\partial R}$, $f_{\mathcal{A}}=\frac{\partial f}{\partial A}$
, $\nabla_{\alpha}$ represents covarient derivative, $\nabla^{2}=\nabla_{\mu}\nabla^{\mu}$ is the d'Alembert operator, and $\mathcal{T}^{\mu\nu}$ implies energy momentum tensor the energy momentum tensor defined as
\begin{equation}
	\mathcal{T}^{\mu\nu}=\frac{1}{\sqrt{-g}} \frac{\delta(\sqrt{-g}\mathcal{L}_{m})}{\delta g_{\mu\nu}}
\end{equation}

\subsection{Linear Model of Ricci Inverse Gravity}
In this research, to keep things simple, we consider the simplest linear model for $f(\mathcal{R}, \mathcal{A})$ gravity defined as \cite{Amendola}
\begin{equation}\label{a6}	
	f(\mathcal{R}, \mathcal{A})=\mathcal{R}+\alpha \mathcal{A},
\end{equation}
where $\alpha$ is the coupling parameter. So in view of Eq. (\ref{a6}), Eq. (\ref{a4}) can be written as
\begin{equation}\label{fieldequation}
	\mathcal{R}^{\mu\nu}-\frac{1}{2}\mathcal{R}g^{\mu\nu}-\alpha \mathcal{A}^{\mu\nu}-\frac{1}{2}\alpha \mathcal{A}g^{\mu\nu}+\frac{\alpha}{2}\bigg[2g^{\gamma\mu}\nabla_{\delta}\nabla_{\gamma}\mathcal{A}^{\delta}_{\zeta}\mathcal{A}^{\nu\zeta}-\nabla^2 \mathcal{A}^{\mu}_{\zeta}\mathcal{A}^{\nu\zeta}-g^{\mu\nu}\nabla_{\delta}\nabla_{\eta}\mathcal{A}^{\delta}_{\zeta}\mathcal{A}^{\eta\zeta}\bigg]=\kappa\mathcal{T}^{\mu\nu}.
\end{equation}
For simplicity, Eq. (\ref{fieldequation}) can be written as
\begin{equation}
	\mathcal{R}^{\mu\nu}-\frac{1}{2}\mathcal{R} g^{\mu\nu}+M^{\mu\nu}=\kappa\mathcal{T}^{\mu\nu},
\end{equation}
where $M^{\mu\nu}$ contains the terms that modify the field equations of general relativity and is given by
\begin{equation}
	M^{\mu\nu}=-\alpha \mathcal{A}^{\mu\nu}-\frac{1}{2}\alpha \mathcal{A}g^{\mu\nu}+\frac{\alpha}{2}\bigg[2g^{\gamma\mu}\nabla_{\delta}\nabla_{\gamma}\mathcal{A}^{\delta}_{\zeta}\mathcal{A}^{\nu\zeta}-\nabla^2 \mathcal{A}^{\mu}_{\zeta}\mathcal{A}^{\nu\zeta}-g^{\mu\nu}\nabla_{\delta}\nabla_{\eta}\mathcal{A}^{\delta}_{\zeta}\mathcal{A}^{\eta\zeta}\bigg],
\end{equation}
the term $M^{\mu\nu}$ arises due to the inclusion of anti-curvature scalar $\mathcal{A}$ in the action defined in Eq. (\ref{a3}).

In addition, the four divergence of the field Eq. (\ref{fieldequation}) yields:
\begin{equation}\label{covarientdivergence}	
	\nabla_{\mu}\mathcal{T}^{\mu\nu}=-\frac{\alpha}{2}g^{\mu\nu}\nabla_{\mu}\mathcal{A}-\alpha\nabla_{\mu}(\mathcal{A}^{\mu\nu})+\alpha g^{\gamma\mu}\nabla_{\mu}\bigg[\nabla_{\delta}\nabla_{\gamma}(\mathcal{A}^{\delta}_\zeta\mathcal{A}^{\nu\zeta})\bigg]-\frac{\alpha}{2}\nabla_{\mu}\bigg[\nabla^{2}(\mathcal{A}^{\mu}_\zeta\mathcal{A}^{\nu\zeta})\bigg]-\frac{\alpha}{2}g^{\mu\nu}\nabla_{\mu}\bigg[\nabla_{\delta}\nabla_{\eta}(\mathcal{A}^{\delta}_\zeta\mathcal{A}^{\eta\zeta})\bigg].
\end{equation}

\section{Interior space-time and Field equations}
The interior space-time of a static spherically symmetric compact object can be characterized by examining the interior line element in the conventional form, defined by:
\begin{equation}\label{spacetime}
	ds^{2}=-e^{2\psi}dt^{2}+e^{2\lambda}dr^{2}+r^{2}d\theta^{2}+r^{2}sin^{2}\theta d\phi^{2},
\end{equation}
where the metric potential functions $\psi$ and $\lambda$ are solely dependent on the radial coordinate $r$. The stress energy-momentum tensor for an anisotropic two-fluid matter configuration can be formulated as follows

	\[\text{Energy-Momentum Tensor    }
	\begin{cases}
		T^{0}_{\mkern8mu 0} = -(\rho + \rho_{q}), \\
		T^{1}_{\mkern8mu 1}  = (p_{r} + p_{q}), \\
		T^{2}_{\mkern8mu 2} =  (p_{t} + p_{q}).
	\end{cases}
	\]
	
The non-zero components of the stress energy-momentum tensor defined above provide information about the anisotropic distribution of matter inside the formation of compact objects. These objects are composed of two forms of matter: ordinary baryonic matter (OBM) and quark matter (QM). In this context, $\rho$ represents the matter-energy density, $p_{r}$ represents the radial pressure component, and $p_{t}$ represents the transverse pressure component of the OBM. Similarly, $\rho_{q}$ represents the matter-energy density, and $p_{q}$ represents the pressure associated with the QM.

By assuming the gravitational units $G=c=1$ and with the use of the non-zero components of the energy-momentum tensor defined above, we can derive a set of the following independent equations as a result of solving Eq. (\ref{fieldequation})
\begin{align}\label{AMJ1}
&8\pi(\rho+\rho_{q})=\mathcal{R}^{00}e^{2\psi}+\frac{\mathcal{R}}{2}-\alpha e^{2\psi}\mathcal{A}^{00}+\frac{\alpha \mathcal{A}}{2}+\frac{\alpha}{2}\bigg[\bigg(\frac{4}{r^{2}}+2\lambda'^{2}+2\lambda''\bigg)(\mathcal{A}^{1}_{\mkern8mu 1}\mathcal{A}^{11})+2re^{-2\lambda}(\mathcal{A}^{2}_{\mkern8mu 2}\mathcal{A}^{22})'-e^{2\psi} e^{-2\lambda}\bigg(\frac{2\psi'}{r}+\notag \\
&\qquad 2\psi'^{2} +2\psi''- 2\lambda' \psi'\bigg)(\mathcal{A}^{0}_{\mkern8mu 0}\mathcal{A}^{00})+3\lambda' (\mathcal{A}^{1}_{\mkern8mu 1}\mathcal{A}^{11})'+(\mathcal{A}^{1}_{\mkern8mu 1}\mathcal{A}^{11})''-e^{2\psi}e^{-2\lambda}(\mathcal{A}^{0}_{\mkern8mu 0}\mathcal{A}^{00})''\bigg],
\end{align}

\begin{align}\label{AMJ2}
&8\pi(p_{r}+p_{q})=\mathcal{R}^{11}e^{2\lambda}-\frac{\mathcal{R}}{2}-\alpha e^{2\lambda}\mathcal{A}^{11}-\frac{\alpha \mathcal{A}}{2}+\frac{\alpha}{2}\bigg[-\bigg(\frac{4\lambda'}{r}+2\lambda' \psi'\bigg)(\mathcal{A}^{1}_{\mkern8mu 1}\mathcal{A}^{11})-2r e^{-2\lambda}(\mathcal{A}^{2}_{\mkern8mu 2}\mathcal{A}^{22})'-4e^{-2\lambda}(\mathcal{A}^{2}_{\mkern8mu 2}\mathcal{A}^{22})\notag \\
&\qquad -\bigg(\frac{2}{r}+\psi'\bigg)(\mathcal{A}^{1}_{\mkern8mu 1}\mathcal{A}^{11})'+\psi' e^{2\psi}e^{-2\lambda}(\mathcal{A}^{0}_{\mkern8mu 0}\mathcal{A}^{00})'+2\psi'^{2}e^{2\psi}e^{-2\lambda}(\mathcal{A}^{0}_{\mkern8mu 0}\mathcal{A}^{00})\bigg],
\end{align}
\begin{align}\label{AMJ3}
&8\pi( p_{t}+p_{q})=r^{2}\mathcal{R}^{22}-\frac{\mathcal{R}}{2}-\alpha r^{2}\mathcal{A}^{22}-\frac{\alpha A}{2}+\frac{\alpha}{2}\bigg[-\bigg(\frac{2}{r^{2}}+2\psi'^{2}+2\lambda'^{2}+2\lambda''\bigg)(\mathcal{A}^{1}_{\mkern8mu 1}\mathcal{A}^{11})+e^{-2\lambda}\bigg(-6r-r^{2}\psi'+r^{2}\lambda'\bigg) \notag \\
&\qquad \times (\mathcal{A}^{2}_{\mkern8mu 2}\mathcal{A}^{22})'+e^{-2\lambda}\bigg(-4+2r\lambda'-2r\psi'\bigg)(\mathcal{A}^{2}_{\mkern8mu 2}\mathcal{A}^{22})+\psi'e^{2\psi}e^{-2\lambda}(\mathcal{A}^{0}_{\mkern8mu 0}\mathcal{A}^{00})'-r^{2}e^{-2\lambda}(\mathcal{A}^{2}_{\mkern8mu 2}\mathcal{A}^{22})''-3\lambda'(\mathcal{A}^{1}_{\mkern8mu 1}\mathcal{A}^{11})'\notag \\
& \qquad -(\mathcal{A}^{1}_{\mkern8mu 1}\mathcal{A}^{11})''\bigg],
\end{align}
where the values of $\mathcal{R}^{00}$, $\mathcal{R}^{11}$, $\mathcal{R}^{22}$, $\mathcal{R}$, $\mathcal{A}^{00}$, $\mathcal{A}^{11}$, $\mathcal{A}^{22}$, $\mathcal{A}$, $\mathcal{A}^{0}_{\mkern8mu 0}\mathcal{A}^{00}$, $\mathcal{A}^{1}_{\mkern8mu 1}\mathcal{A}^{11}$, and $\mathcal{A}^{2}_{\mkern8mu 2}\mathcal{A}^{22}$ are given in the appendix section.

To solve the above set of Eqs. (\ref{AMJ1})-(\ref{AMJ3}) for normal baryonic matter, we have to close the system. For this purpose we can use the relationship between matter density $\rho$ and radial pressure $p_{r}$ defined as
\begin{equation}\label{EoS1}
	p_{r}=\beta_{1} \rho-\beta,
\end{equation}
where $\beta_{1}$ and $\beta$ are positive constants and $\beta_{1}$ lies in the range $0<\beta_{1}<1$ with $\beta_{1}\neq \frac{1}{3}$. Let us also assume that, for QM, the relationship between pressure $p_{q}$ and matter density $\rho_{q}$ is described by the following MIT Bag EoS model \cite{MITEoS}
\begin{equation}\label{EoS2}
	p_{q}=\frac{1}{3}(\rho_{q}-4B),
\end{equation}
where $B$ describes the Bag constant.
\section{Solution of Field Equations}
In order to determine the physical attributes of our model, we have to solve field  Eqs. (\ref{AMJ1})-(\ref{AMJ3}) along with the EoS describe in Eq. (\ref{EoS1}) and Eq. (\ref{EoS2}). For this, we can Utilize the parametrization for the metric potentials proposed by Krori and Barua \cite{PR54} (known as KB ansatz) which is defined as:

	\[\text{KB metric     }
\begin{cases}
	e^{2\psi(r)} = e^{Hr^{2}+P}, \\
	e^{2\lambda(r)}=e^{Gr^{2}}, \\
\end{cases}
\]

here, $G$ and $H$ are constant parameters measured in the units of $km^{-2}$, with $P$ being a dimensionless constant. These constants can be evaluated numerically by using the smooth matching of interior and exterior space-times. We are primarily motivated to introduce this collection of functions for the static metric goes back to the singularity free structure within the star's boundary.

With the choice of linear model for Ricci inverse gravity defined in Eq. (\ref{a6}) together with the choice of KB metric potentials and the EoSs for both OBM and QM defined in Eqs. (\ref{EoS1})-(\ref{EoS2}), the solution of the Eqs. (\ref{AMJ1})-({\ref{AMJ3}}) yields the values of $\rho$, $p_{r}$, $p_{t}$, $\rho_{q}$, and $p_{q}$, for both OBM and QM. The mathematical expressions are given in the appendix section.
\subsection{Regularity of our chosen metric}
This subsection focuses on the behavior of the KB metric potentials that we have chosen. We note that $e^{2\psi}|_{r=0}=e^{P}> 0$ and $e^{2\lambda}|_{r=0}= 1$, which identifies that both metric potentials are finite at the center and have regularity throughout the domain $r<R$. Moreover, the derivatives of the metric coefficients are: $\big[\frac{d}{dr}(e^{2\lambda})\big]_{r=0} = 2Gre^{Gr^{2}}|_{r=0}=0$ and $\big[\frac{d}{dr}(e^{2\psi})\big]_{r=0} = 2Hre^{Hr^{2}+P}|_{r=0}=0$. These values define that, at the core of the star, the derivatives of the metric potentials are equal to zero. Hence, the selection of the metric potentials in this context guarantees that the metric functions exhibit non-singularity and continuity within the star's interior, and these metric potentials are perfectly aligned to the external Schwarzschild line element. On a physical basis, this is considered a desirable characteristic for any well-behaved model.
\subsection{Calculation of Metric Potentials from Junction Conditions}
To solve the field equations outlined in Eq.(\ref{AMJ1})-({\ref{AMJ3}}), we establish the values of the parameters $G$, $H$, and $P$ appearing in the KB metric. For this purpose, we match the interior spacetime defined in Eq. (\ref{spacetime}) to the exterior spacetime under the restricted boundary conditions at $r=R$. The exterior metric of the star, which is described by the following Schwarzschild solution given by
\begin{equation}
ds^{2}=-\bigg(1-\frac{2M}{r}\bigg)dt^2+\bigg(1-\frac{2M}{r}\bigg)^{-1}dr^{2}+r^{2}d\theta^{2}+r^{2}sin^{2}\theta d\phi^{2},
\end{equation}
here, $M$ represents the total mass enclosed within the boundary of the compact star. Ensuring the continuity of the metric coefficients $g_{tt}$, $g_{rr}$, and $\frac{\partial g_{tt}}{\partial r}$ across the boundary surface $r = R$, separating the interior and exterior regions yields the following set of relationships:

\begin{equation}\label{continuity1}
	1-\frac{2M}{R}=e^{Hr^{2}+P},
\end{equation}
\begin{equation}\label{continuity2}
	\bigg(1-\frac{2M}{R}\bigg)^{-1}=e^{Gr^{2}},
\end{equation}
\begin{equation}\label{continuity3}
	\frac{M}{R^{2}}=HRe^{HR^{2}+P}.
\end{equation}
Upon solving the Eqs. (\ref{continuity1})-(\ref{continuity3}) yields the following values of constants $G$, $H$, and $P$ in terms of the total mass $M$ and radius $R$, and are given by
\begin{equation}
G=-\frac{1}{R^{2}}ln\bigg(1-\frac{2M}{R}\bigg),
\end{equation}
\begin{equation}
H=\frac{M}{R^{3}}\bigg(1-\frac{2M}{R}\bigg)^{-1},
\end{equation}
\begin{equation}
P=ln\bigg(1-\frac{2M}{R}\bigg)-\frac{M}{R}\bigg(1-\frac{2M}{R}\bigg)^{-1}.
\end{equation}
Furthermore, at star's boundary where radius of star is indicated as $R$, the pressure in radial direction vanishes, denoted mathematically as $p_{r}(r=R)=0$. This condition leads to the expression for $\beta$ as provided in the appendix.

As a result, we have successfully determined the values of $G$, $H$, and $P$ involved in the KB metric coefficients in terms of mass $M$ and radius $R$. Using the observed values of various candidates of compact stars, we can find the numerical values of the constants $G$, $H$, and $P$.

\section{Physical attributes of present model}
In this section, we graphically analyze the physical attributes of our present hybrid star model. Focusing especially on the star LMC-X4 with observed mass $M = (1.04^{+0.09}_{-0.09})M_{\odot}$ and observed radius $R=8.301^{+0.2}_{-0.2} \text{ km}$. The entire analysis has been conducted with the choice of $\beta_{1}=0.3$.

\subsection{Nature of fluid components associated with OBM and QM}
\begin{figure}
	\centering \epsfig{file=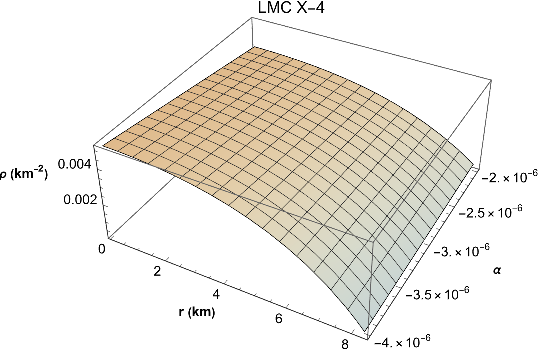, width=.40\linewidth,height=1.70in}
	\centering \epsfig{file=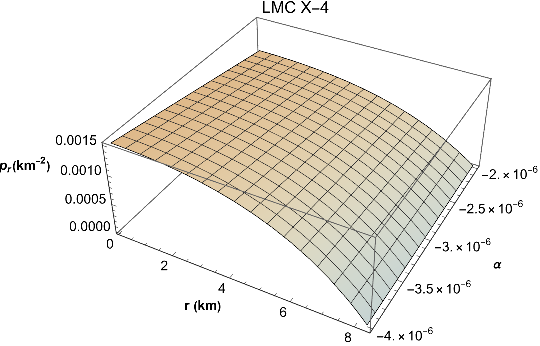, width=.40\linewidth,height=1.70in}
	\centering \epsfig{file=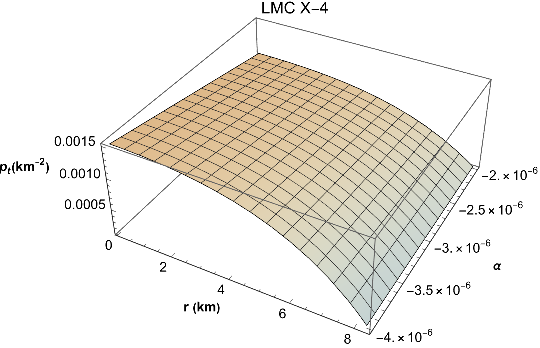, width=.40\linewidth,height=1.70in}
	\centering \epsfig{file=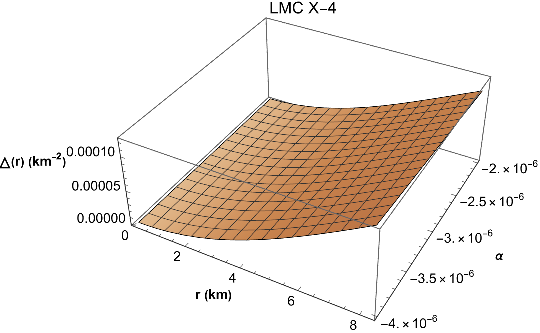, width=.40\linewidth,height=1.70in}
	\caption{\label{Fig.1a} Profile of baryonic matter-energy density $\rho$, pressure components $(p_{r}, p_{t})$, and anisotropic factor $\Delta(r)$ against $r$.}
\end{figure}

\begin{figure}
	\centering \epsfig{file=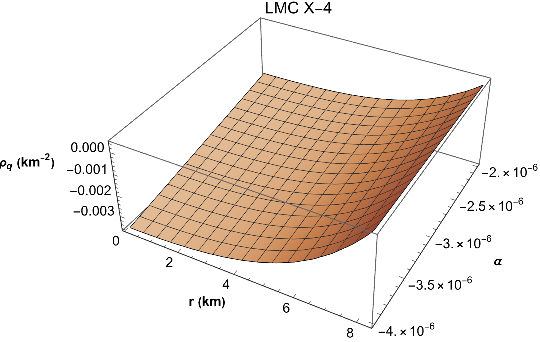, width=.40\linewidth,height=1.70in}
	\centering \epsfig{file=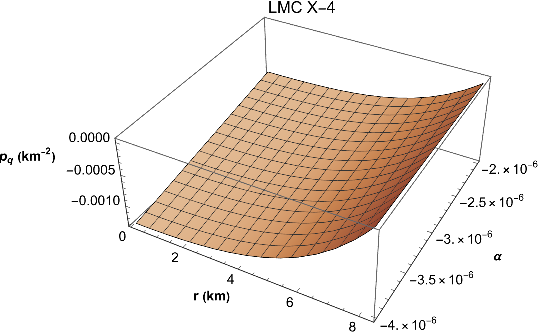, width=.40\linewidth,height=1.70in}
	\caption{\label{Fig.2a} Profile of QM density $\rho_{q}$ and QM pressure $p_{q}$ against $r$.}
\end{figure}

In this subsection, we examine the behaviors of energy density $\rho$, radial pressure $p_{r}$, and tangential pressure $p_{t}$ in the context of outer baryonic matter OBM, as well as $\rho_{q}$ and $p_{q}$ for the QM. Additionally, we explore the gradients of $\rho$, $p_{r}$, and $p_{t}$. Due to the very complicated analytic forms, we will examine their behaviors through the graphical representations. In Fig. (\ref{Fig.1a}), we have plotted the OBM matter-energy density and pressure components. The graphical analysis shows the consistent decreasing behaviors of $\rho$, $p_{r}$, and $p_{t}$ against radius $r$ for different values of coupling parameter $\alpha$, as shown in the figure. The graphical visualization shows that $\rho$, $p_{r}$, and $p_{t}$ have the maximum value at the core of the star. Also, $\rho$, $p_{r}$, and $p_{t}$ are non-negative inside the star. As anisotropic matter distribution has been taken into consideration, indicating a difference between pressure components $(p_{r}, p_{t})$. Therefore, $\Delta=p_{t}-p_{r}$ is used to define the anisotropy factor. The anisotropic force, denoted by $\frac{2\Delta}{r}$, will be replusive in nature if $\Delta>0$ and attractive otherwise. The bottom right panel of Fig. (\ref{Fig.1a}) shows the profile for the anisotropy factor.
In Fig. (\ref{Fig.2a}), we have graphed the energy density $\rho_{q}$ and pressure $p_{q}$ for the QM . This Fig. shows that both display a negative nature inside the compact stellar object.

\begin{figure}
	\centering \epsfig{file=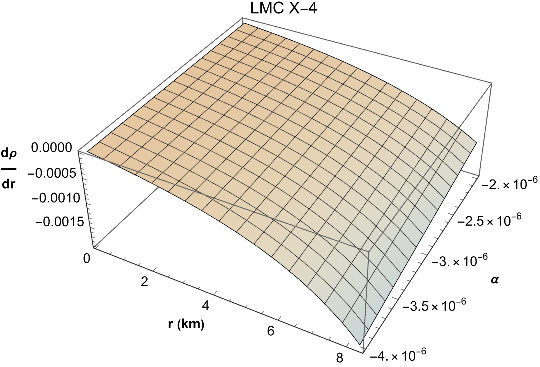, width=.40\linewidth,height=1.70in}
	\centering \epsfig{file=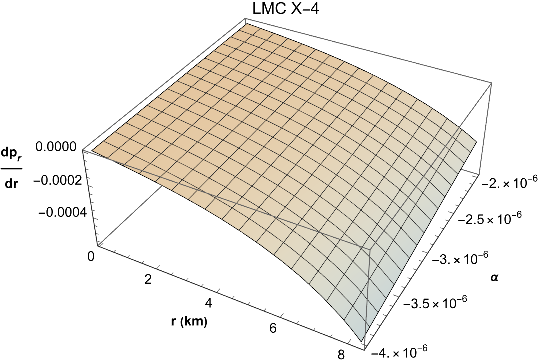, width=.40\linewidth,height=1.70in}
	\centering \epsfig{file=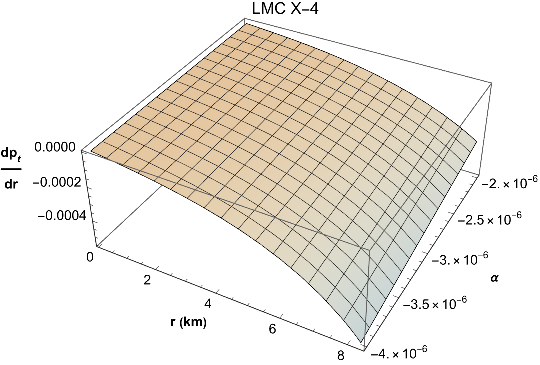, width=.40\linewidth,height=1.70in}
	\caption{\label{Fig.3a} Profiles of $\frac{d\rho}{dr}$, $\frac{dp_{r}}{dr}$, and $\frac{dp_{t}}{dr}$ against $r$ inside the stellar interior.}
\end{figure}

In our present model, Fig. (\ref{Fig.3a}) shows the variation of density and pressure gradients due to OBM. The graphical illustration shows that the gradient of density and pressure due to OBM stays negative throughout the fluid sphere, which is expected for a realistic physical model.

\subsection{Energy Conditions}
In order to ensure the physical realism and feasibility of the matter field, it is necessary for the stress-energy tensor to adhere to certain mathematical contstraints. We generally referred to these constraints as energy conditions (Ecs). The analysis of ECs is essential in comprehending the properties of ordinary and unusual nature of matter within the stellar structure. The ECs are depicted as follows \cite{hybrid_star_fG_PR_90, hybrid_star_fG_PR_91, hybrid_star_fG_PR_92, hybrid_star_fG_PR_93, hybrid_star_fG_PR_94}:
 \\
\begin{itemize}
	\item Null Energy Condition (NEC)~~~~~~~~~~~~~$\rho+p_{i}\geq 0$,
	
	\item Weak Energy Condition (WEC)~~~~~~~~~~~$\rho\geq0$,~~~$\rho+p_{i}\geq 0$,
	\item Strong Energy Condition (SEC)~~~~~~~~~~~$\rho+p_{i}\geq 0$,~~~$\rho+p_{r}+2p_{t}\geq 0$,
	\item Dominant Energy Condition (DEC)~~~~~~$\rho-p_{i}>0$,~~~ $\rho\geq0$
\end{itemize}
where $i=r,t$.

\begin{figure}
	\centering \epsfig{file=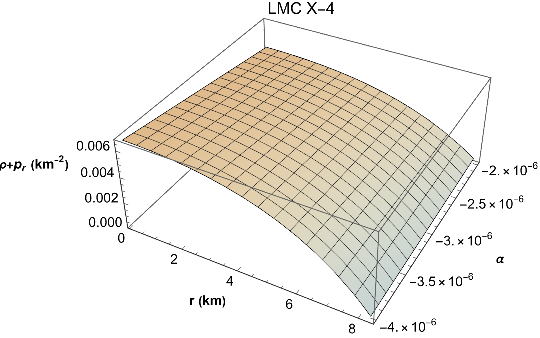, width=.40\linewidth,height=1.70in}
	\centering \epsfig{file=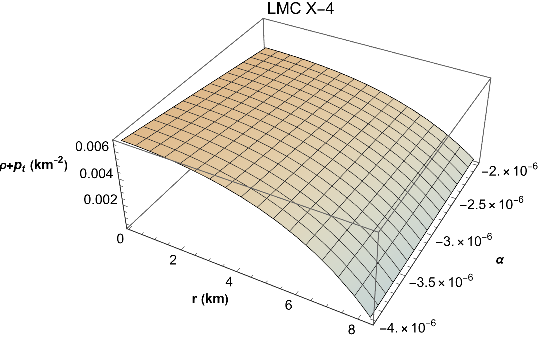, width=.40\linewidth,height=1.70in}
	\centering \epsfig{file=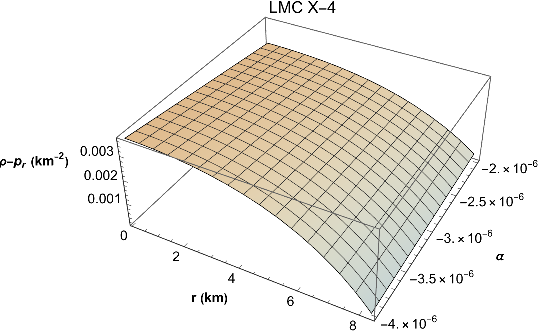, width=.40\linewidth,height=1.70in}
	\centering \epsfig{file=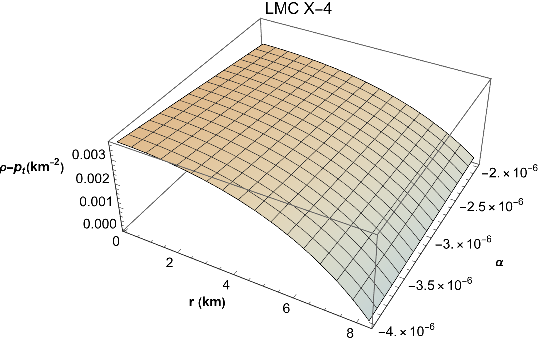, width=.40\linewidth,height=1.70in}
	\centering \epsfig{file=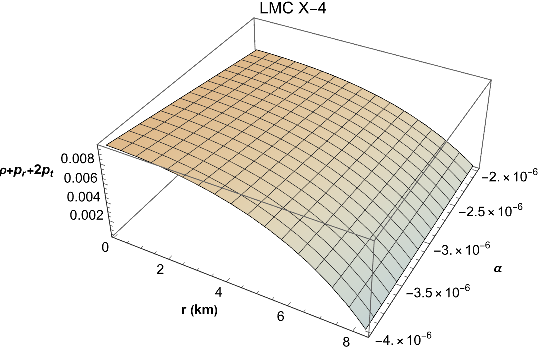, width=.40\linewidth,height=1.70in}
	\caption{\label{Fig.8a} Behavior of all energy conditions against $r$ inside the stellar interior for LMC-X4.}
\end{figure}

Ordinary matter always satisfies these ECs due to its inherent positive energy density and pressure. Fig. (\ref{Fig.8a}) shows the graphical visualization of $\rho+{p_{r}}$, $\rho+{p_{t}}$, $\rho-{p_{r}}$, $\rho-{p_{t}}$, and $\rho+{p_{r}}+2p_{t}$ for different values of $\alpha$, and this Fig. visually confirm the validity of the ECs. The ECs mentioned earlier are all met by the proposed hybrid star model in Ricci Inverse gravity suggesting our model is physically acceptable.

\subsection{Equation of State}
\begin{figure}
	\centering \epsfig{file=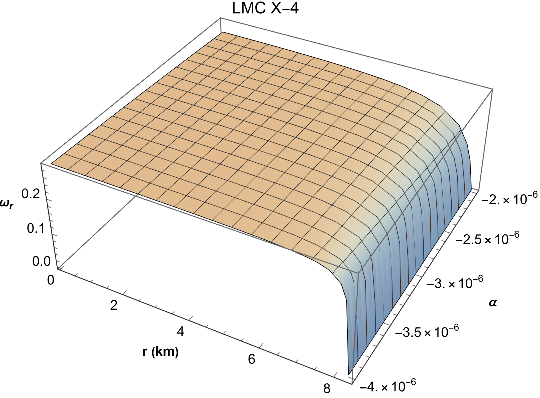, width=.40\linewidth,height=1.70in}
	\centering \epsfig{file=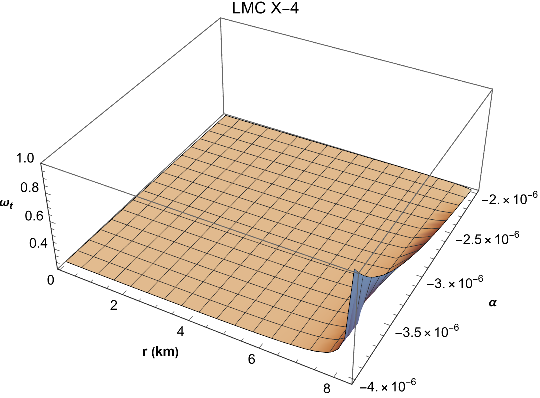, width=.40\linewidth,height=1.70in}
	\caption{\label{Fig.5a} Profiles for $w_{r}$ and $w_{t}$ against $r$.}
\end{figure}

The equation of state (EoS) is a fundamental relation that establishes the relationship between radial pressure and energy density, as well as tangential pressure and energy density. Creating two separate EoSs is advantageous, with one dedicated to radial pressure and the other to tangential pressure. In our study, we establish the EoS by employing the following equations:

\begin{equation}
	p_{r}=\omega_{r} \rho, ~~~~~ p_{t}=\omega_{t} \rho,
\end{equation}
The two dimensionless parameters, denoted as $\omega_r$ and $\omega_t$, are commonly referred to as EoS parameters that describe the relation between matter density and pressure, and these parameters also give information about the state of matter. The graphical profiles of $\omega_r$ and $\omega_t$ are depicted in Fig. (\ref{Fig.5a}) for different variations in $\alpha$.  The findings indicate that the two features were most beneficial closer to the star's center and fell within the ranges $0<\omega_{r}<1$ and $0<\omega_{t}<1$.

\subsection{MASS FUNCTION, COMPACTNESS FACTOR, SURFACE REDSHIFT}
\begin{figure}
	\centering \epsfig{file=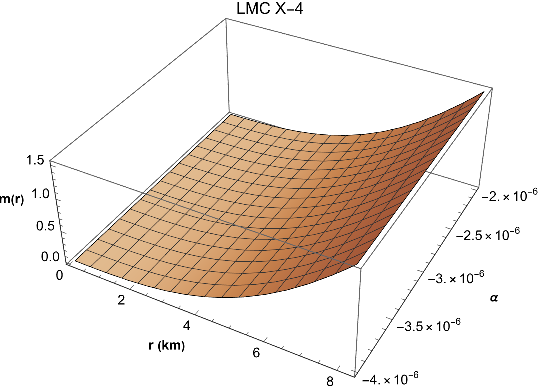, width=.40\linewidth,height=1.70in}
	\centering \epsfig{file=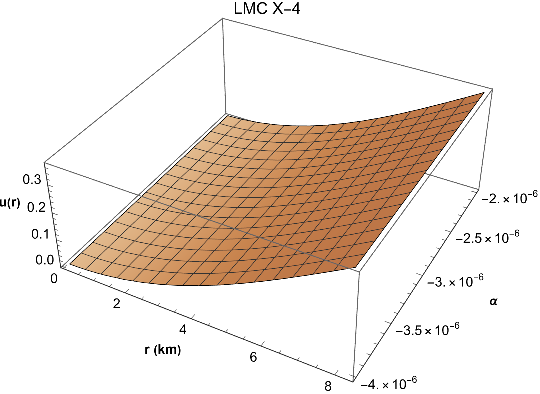, width=.40\linewidth,height=1.70in}
	\centering \epsfig{file=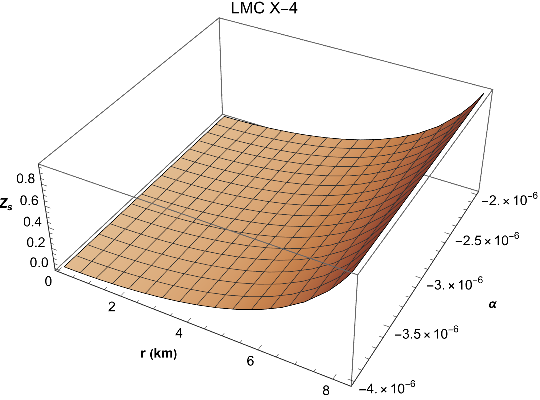, width=.40\linewidth,height=1.70in}
	\caption{\label{Fig.6a} Variations of mass function $m(r)$, compactness factor $u(r)$, and surafce redshift $Z_{s}$ are shown against radius $r$.}
\end{figure}

In this subsection, we will look at some key parameters of our model: mass function $m(r)$, compactness factor $u(r)$, and surface redshift $Z_{s}$. The function $m(r)$ plays a significant role as it represents the distribution of mass within the star as a function of radial distance $r$ and is defined by the following formula:
\begin{equation}\label{mass}
	m(r)=\int_{0}^{r} 4\pi (\rho+\rho_{q})\zeta^{2}d\zeta.
\end{equation}
after applying the KB potentials on Eq. (\ref{mass}), we finally get \cite{hybrid_star_fG_PR_86, hybrid_star_fG_PR_87}
\begin{equation}
	m(r)=\frac{r}{2e^{2\lambda}}\bigg(-1+e^{2\lambda}\bigg)=\frac{r}{2}\bigg(1-e^{-Gr^{2}}\bigg)
\end{equation}
Fig. (\ref{Fig.6a}) displays the variation of $m(r)$ against $r$. It is evident from the Fig. that there is no singularity in the function $m(r)$ and possesses monotonically increasing behavior and having maximum value at the surface of the star. Also graphical illustration shows that when $r\rightarrow 0$, $m(r)\rightarrow 0$.
The compactness factor $u(r)$ plays a crucial role in understanding the gravitational properties and structural stability of these dense astronomical objects. It is defined as the ratio of mass of the object to its radius. Mathematically,
\begin{equation}
	u(r)=\frac{m(r)}{r}
\end{equation}
The graphical evolution of $u(r)$ has been displayed in Fig. (\ref{Fig.6a}) and shows the monotonically increasing behavior against $r$ and having maximum value at the surface of the star. The surface redshift $Z_{s}$ for the present compact star candidate can be obtained by using the following expression:
\begin{equation}
	Z_{s}=(1-2u(r))^{-\frac{1}{2}}-1
\end{equation}
The graphical evolution of $Z_{s}$ for different values of $\alpha$ is also shown in Fig. (\ref{Fig.6a}). This graph shows the increasing behavior of $Z_{s}$ and its maximum value at the star's surface.

\section{Stability Analysis}
In this section, we will evaluate the stability of our model by examining the causality condition, adiabatic index, and equilibrium condition.

\subsection{Velocity of sound}
Analyzing sound velocity can provide valuable information about the properties of dense matter in celestial bodies like hybrid stars or quark stars.
The speed of sound within a stellar object is a crucial determinant that impacts its overall stability. In order to establish a physically accurate model, it is imperative to verify the causality condition, which implies that the velocity of sound within the compact object must be less than the speed of light. The subsequent formulas for the radial sound speed associated with the radial pressure denoted by $v_{r}^{2}$ and tangential sound speed denoted by $v_{t}^{2}$ can be utilized, which are expressed as.

\begin{equation}
	v_{r}^{2}=\frac{dp_{r}}{d\rho}, ~~~v_{t}^{2}=\frac{dp_{t}}{d\rho}
\end{equation}
The stability of a viable system is significantly shaped by the concept introduced by Herrera \cite{Amina91}. Based on this theoretical work, the speed of sound associated to both the radial and transverse pressure components must adhere to the causality condition, where the speed of sound falls within the range of [0,1] i.e., $0 \leq v_{r}^2 \leq 1$ and $0 \leq v_{t}^2 \leq 1$. Abreu \cite{Amina92} introduced a significant concept for examining potential stable and unstable configurations of celestial objects. The constraints imposed on $v_{r}^2$ and $v_{t}^2$ lead to the following condition:

	\[
	\begin{cases}
		-1 \leq v_{t}^2-v_{r}^2\leq0 ~~~~ \text{indicating potential stability},\\
		~~0 \leq v_{t}^2-v_{r}^2\leq1 ~~~~ \text{indicating potential instability}.
	\end{cases}
	\]
	
\begin{figure}
	\centering \epsfig{file=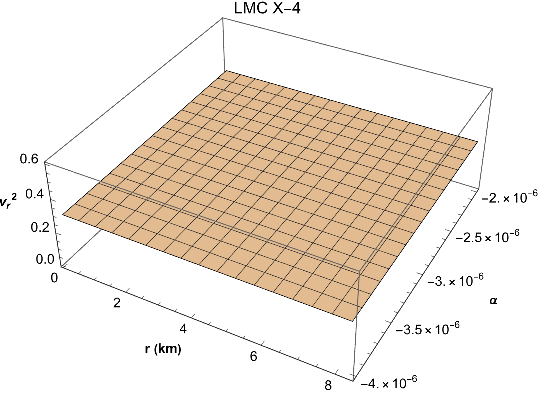, width=.40\linewidth,height=1.70in}
	\centering \epsfig{file=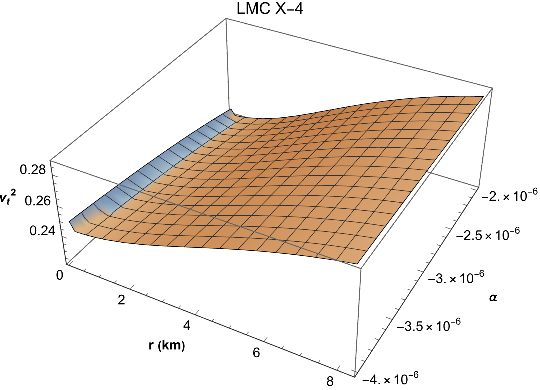, width=.40\linewidth,height=1.70in}
	
	\caption{\label{Fig.4a} Plotting of squares of radial sound velocity $v_{r}^2$ and transverse sound velocity $v_{t}^{2}$ against $r$.}
\end{figure}

Due to the complexities in the expressions of $v_{r}^{2}$ and $v_{t}^{2}$, we analyzed these terms graphically. Fig. (\ref{Fig.4a}) shows the graphical behavior of both $v_{r}^{2}$ and $v_{t}^{2}$ , revealing that they are both restricted within the range $[0,1]$ within the boundary of stellar object. This condition is commonly referred as the causality condition.

\subsection{Adiabatic Index}
\begin{figure}
	\centering \epsfig{file=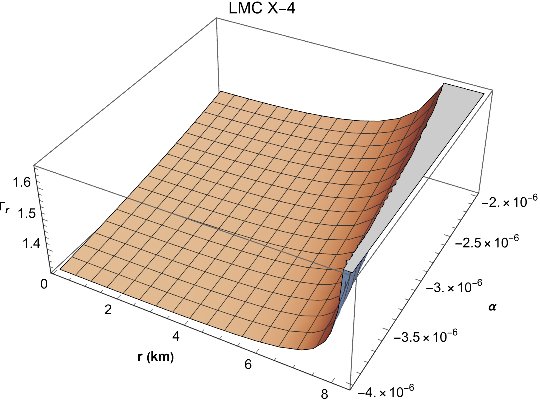, width=.40\linewidth,height=1.70in}
	\centering \epsfig{file=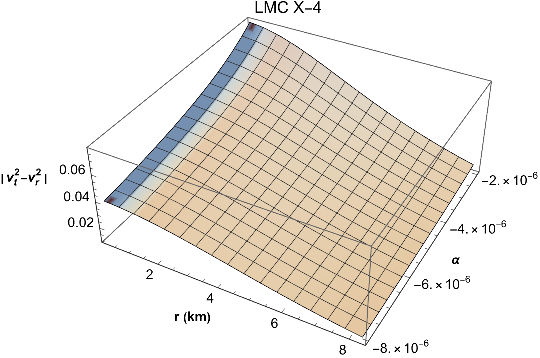, width=.40\linewidth,height=1.70in}
	\caption{\label{Fig.7a} Behavior of $\Gamma_{r}$ and stability factor $|v_{t}^{2}-v_{r}^{2}|$ against $r$ .}
\end{figure}

Here, we delve into the analysis of a crucial ratio defined as the adiabatic index denoted by $\Gamma$ to assess the stability region of the hybrid star model.
Chen et al. \cite{adnan102} introduced the concept of adiabatic index for an isotropic fluid sphere, however, Chandrasekhar \cite{adnan103} was one of the pioneers in utilizing the adiabatic index to examine the stability zone of a spherical star. The following expression can be used for the adiabatic index in the presence of anisotropy in the pressure:
\begin{equation}
	\Gamma_{r}=\frac{\rho+p_{r}}{p_{r}}\frac{dp_{r}}{d\rho},
\end{equation}
where $\Gamma_{r}$ is the adiabatic index for radial pressure.
According to the study conducted by Heintzmann and Hillebtanddt \cite{adnan104}, the stellar object satisfies the conditions of stability when the values of the aforementioned expression $\Gamma_{r}$ exceed 4/3. Due to the intricate nature of the expressions, it is unfeasible to analytically prove this requirement. The profiles of $\Gamma_r$ for different values of $\alpha$ have been plotted for our model, and we see that $\Gamma_r$ is an increasing function of $r$ and $\Gamma_r>\frac{4}{3}$ everywhere inside the boundary of the star as depicted in Fig. (\ref{Fig.7a}). The graph illustrates that $\Gamma_r$ exhibits values exceeding 4/3 throughout the fluid sphere, confirming the complete satisfaction of the stability condition. The right panel of Fig. (\ref{Fig.7a}) shows the graphical illustration of $|v_{t}^{2}-v_{r}^{2}|$, confirming the potential stability of our considered Ricci Inverse gravity.

\subsection{Equilibrium Condition}
In this subsection, we will look at the model's equilibrium with respect to various forces acting on the system. The equilibrium equation involves four main forces: hydrostatic force $(F_{h})$, gravitational force $(F_{g})$, anisotropic force $(F_{a})$, and the force associated with quark matter $(F_{q})$. Additionally, there are some extra terms arises due to the modification the gravitational theory and we have labeled these terms as $F_{\text{ext}}$.

The Tolman-Oppenheimer-Volkoff (TOV) equation serves as a crucial framework for comprehending this equilibrium.
To assess the stability of our model, we formulate the modified version of the TOV equation within the context of the $f(\mathcal{R}, \mathcal{A})$ gravity theory. So Eq. (\ref{covarientdivergence}) yields the following equation
\begin{align}\label{a140}
	& -\frac{dp_{r}}{dr}-\bigg(\rho+p_{r}\bigg)\psi'+ \frac{2\Delta}{r}+\alpha\bigg[-\bigg(\psi'+2\lambda'+\frac{2}{r}\bigg)e^{2\lambda}\mathcal{A}^{11}-\psi' e^{2\psi}\mathcal{A}^{00}+2r\mathcal{A}^{22}\bigg]-\frac{\alpha}{2}(\mathcal{A})'-\alpha e^{2\lambda}(\mathcal{A}^{11})'+ \notag \\
	&\qquad \frac{\alpha}{2r^3}\bigg[\bigg(-10r+8r^2\lambda'+4r^3\lambda'\psi'-6r^3\psi'^2-r^3\psi''\bigg)(\mathcal{A}^{1}_{\mkern8mu 1}\mathcal{A}^{11})'+\bigg(2r^2+2r^3\psi'\bigg)(\mathcal{A}^{1}_{\mkern8mu 1}\mathcal{A}^{11})''+e^{-2\lambda}\bigg(2r^3+4r^4\lambda'\bigg)\notag \\
	& \qquad \times (\mathcal{A}^{2}_{\mkern8mu 2}\mathcal{A}^{22})'-2r^{4}e^{-2\lambda}(\mathcal{A}^{2}_{\mkern8mu 2}\mathcal{A}^{22})''+r^{3}\psi'e^{-2\lambda}(\mathcal{A}^{0}_{\mkern8mu 0}\mathcal{A}^{00})''+\bigg(-2r^{3}\psi'\lambda'+r^3\psi''\bigg)(\mathcal{A}^{0}_{\mkern8mu 0}\mathcal{A}^{00})'+\bigg(16+8r^2\lambda'^2+\notag \\
	& \qquad 4r^{3}\psi'\lambda'^2+4r^{2}\lambda''+ 4r^3\psi'^3+2r^3\psi'\lambda''-4r^3\psi'\psi''-12r\lambda'-8r^{3}\lambda'\psi'^2-2r^3\lambda'\psi''\bigg)(\mathcal{A}^{1}_{\mkern8mu 1}\mathcal{A}^{11})\bigg]-(\rho_{q}+p_{q})\frac{\psi'}{2}-\notag \\
	& \qquad \frac{d}{dr}(p_{q})=0,
\end{align}
here, prime indicates the derivative concerning the radial coordinate $r$.
The above equation can be viewed as
\begin{equation}\label{a1400}
F_{h}+F_{g}+F_{a}+F_{\text{ext}}+F_{q}=0.
\end{equation}
 Eq. (\ref{a140}) establishes the equilibrium condition of the configuration, considering the mentioned forces, which can be expressed as follows:
 \begin{align}\label{a1400}
 	F_{h}=-\frac{dp}{dr},
 \end{align}
 \begin{align}\label{a1400}
 	F_{a}=\frac{2}{r}(p_{t}-p_{r})=\frac{2}{r}\Delta,
 \end{align}
 \begin{align}\label{a1400}
 	F_{g}=	& -\bigg(\rho+p_{r}\bigg)\psi'
 \end{align}
 \begin{align}\label{a1400}
 	F_{q}=	& -\bigg(\rho_{q}+p_{q}\bigg)\frac{\psi'}{2}-\frac{d}{dr}(p_{q})
 \end{align}
  \begin{align}\label{t1400}
 	F_{\text{ext}}=	& \alpha\bigg[-\bigg(\psi'+2\lambda'+\frac{2}{r}\bigg)e^{2\lambda}\mathcal{A}^{11}-\psi' e^{2\psi}\mathcal{A}^{00}+2r\mathcal{A}^{22}\bigg]-\frac{\alpha}{2}(\mathcal{A})'-\alpha e^{2\lambda}(\mathcal{A}^{11})'+ \notag \\
 	&\qquad \frac{\alpha}{2r^3}\bigg[\bigg(-10r+8r^2\lambda'+4r^3\lambda'\psi'-6r^3\psi'^2-r^3\psi''\bigg)(\mathcal{A}^{1}_{\mkern8mu 1}\mathcal{A}^{11})'+\bigg(2r^2+2r^3\psi'\bigg)(\mathcal{A}^{1}_{\mkern8mu 1}\mathcal{A}^{11})''+e^{-2\lambda}\bigg(2r^3+\notag \\
 	& \qquad 4r^4\lambda'\bigg) (\mathcal{A}^{2}_{\mkern8mu 2}\mathcal{A}^{22})'-2r^{4}e^{-2\lambda}(\mathcal{A}^{2}_{\mkern8mu 2}\mathcal{A}^{22})''+r^{3}\psi'e^{-2\lambda}(\mathcal{A}^{0}_{\mkern8mu 0}\mathcal{A}^{00})''+\bigg(-2r^{3}\psi'\lambda'+r^3\psi''\bigg)(\mathcal{A}^{0}_{\mkern8mu 0}\mathcal{A}^{00})'+\notag \\
 	& \qquad \bigg(16+8r^2\lambda'^2+ 4r^{3}\psi'\lambda'^2+4r^{2}\lambda''+ 4r^3\psi'^3+2r^3\psi'\lambda''-4r^3\psi'\psi''-12r\lambda'-8r^{3}\lambda'\psi'^2-2r^3\lambda'\psi''\bigg)(\mathcal{A}^{1}_{\mkern8mu 1}\mathcal{A}^{11})\bigg].
 \end{align}

\begin{figure}
	\centering \epsfig{file=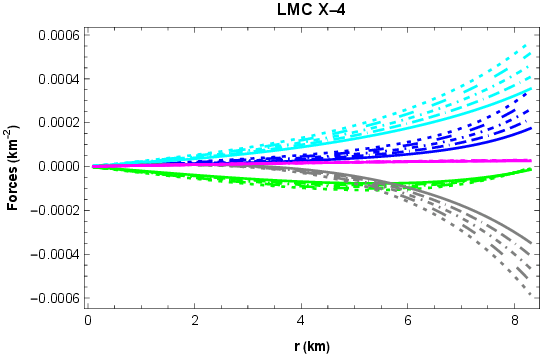, width=.45\linewidth,height=1.95in}
	\caption{\label{Fig.10a} Variations of $\protect\textcolor{green}{\bigstar}$ gravitational, $\protect\textcolor{cyan}{\bigstar}$ hydrostatic, $\protect\textcolor{magenta}{\bigstar}$ anisotropic, $\protect\textcolor{gray}{\bigstar}$ quark, and $\protect\textcolor{blue}{\bigstar}$ $F_{\text{ext}}$ forces against $r$ for $\alpha=-2\times 10^{-6}$, $\alpha=-2.5\times 10^{-6}$, $\alpha=-3\times 10^{-6}$, $\alpha=-3.5\times 10^{-6}$, and $\alpha=-4\times 10^{-6}$.}
\end{figure}

The Fig. (\ref{Fig.10a}) illustrates the varying behaviors of all the forces $F_{g}$, $F_{h}$, $F_{a}$, $F_{\text{ext}}$, and $F_{q}$ acting on the model under concern for different variations in $\alpha$. This figure also confirms that the collective impact of all forces ensures the stability of our model.


\section{CONCLUDING REMARKS}
The challenge of determining an appropriate model to depict the intricate geometry of compact objects has garnered attention in the field of GR and extended theories of gravity. In this investigation, we have developed a method to examine the potential development of a hybrid star model within the framework of modified Ricci inverse gravity $f(\mathcal{R}, \mathcal{A})$, an extension to GR.
 Modeling such astrophysical objects poses significant challenges using $f(\mathcal{R},\mathcal{A})$. Due to the complexities of the field equation of this modified gravity, we opted for the linear model: $f(\mathcal{R},\mathcal{A})=\mathcal{R}+\alpha \mathcal{A}$. The stellar model has been developed based on the choice of metric potentials known as Krori Barua metric potentials that involve the arbitrary constant $G$, $H$, and $P$. For our analysis, we have considered the compact star LMC-X4, and all investigation has been done by considering a hybrid stellar model with an anisotropic matter distribution comprising both OBM and QM. By using the observed values of mass and radius of this star, we have calculated the values of the aforementioned arbitrary constants by matching interior space-time with Schwarzschild space-time. The physical analysis of our results reveals that this anisotropic hybrid stellar model in $f(\mathcal{R},\mathcal{A})$ gravity exhibits the following conclusive properties:

\begin{itemize}

\item The chosen KB metric demonstrates a consistent and well-behaved description of the gravitational field by maintaining regularity throughout the stellar interior. The absence of singularities within the stellar interior underscores the reliability of the KB metric in characterizing gravitational dynamics.

\item  The behavior of $\rho$ and pressure components $(p_{r}, p_{t})$ corresponding to the OBM is positive and decreasing across the interior domain of the star and having maximum values at the core as depicted in Fig. (\ref{Fig.1a}).

\item Considering the presence of anisotropic matter, the difference between $p_{t}$ and $p_{r}$, which is denoted by $\Delta$, is non-zero. The bottom right panel of Fig.  (\ref{Fig.1a}) visually represents the behavior of $\Delta$. Its positive behavior signifies that anisotropic force is regenerative in nature.

\item The energy density $\rho_{q}$ and pressure profiles $p_{q}$ exhibit negative behaviors due to QM within the compact stellar object, as demonstrated in Fig. (\ref{Fig.2a}).
\item The pressure and density gradients corresponding to OBM remain negative throughout the fluid sphere, which one would anticipate from a physically accurate model.

\item The mass function $m(r)$ is directly proportional to the radial distance $r$ and exhibits positive and increasing behavior in the interior of the star, having maximum value at the surface.

\item The compactness factor and surface redshift exhibit a continuous increase as the radial distance $r$ grows.

\item The causality condition is satisfied as both radial and transverse sound speeds, represented as $v_{r}^{2}$ and $v_{t}^{2}$ respectively, remain within the bound $[0, 1]$ throughout the interior of the stellar object. Furthermore, our model adheres to Herrera's cracking criterion, indicating its physical consistency and potential stability throughout the stellar distribution in $f(\mathcal{R},\mathcal{A})$ gravity.

\item All ECs show positive behavior throughout the interior of the star, showing the presence of ordinary matter, so this proves the existence a realistic matter content.

\end{itemize}
Various stellar solutions have been derived by using diverse modified theories of gravity, confirming the reliability of these types of frameworks. As a result of all the notable findings, it is evident that we can build a viable, stable, and singularity-free generalized hybrid stellar model within the interior fluid distribution in this specific $f(\mathcal{R},\mathcal{A})$ gravity framework. Hence, it is possible to analyze the physical characteristics of strange star entities using theoretical and astrophysical methods through the study of an exceptionally dense and compact celestial body consisting of QM. All the analysis leads to the conclusion that the hybrid star model presented in this work adequately clarifies the physical characteristics in the framework of Ricci inverse gravity.

\section{Appendix}
\begin{equation}\notag
	\mathcal{R}^{00}=\Big(4 H r^2 (H-G)+6 H\Big) e^{-2 \left(r^2 (G+H)+P\right)}
\end{equation}
\begin{equation}\notag
	\mathcal{R}^{11}=e^{-4 G r^2} \Big(4 H r^2 (G-H)+4 G-2 H\Big)
\end{equation}
\begin{equation}\notag
	\mathcal{R}^{22}=\frac{e^{-2 G r^2} \left(2 r^2 (G-H)-1\right)+1}{r^4}
\end{equation}
\begin{equation}\notag
	\mathcal{R}=\frac{2 e^{-2 G r^2} \left(4 H r^4 (G-H)+2 r^2 (2 G-3 H)+e^{2 G r^2}-1\right)}{r^2}
\end{equation}	
\begin{equation}\notag
	\mathcal{A}^{00}=\frac{e^{-2 \left(r^2 (H-G)+P\right)}}{2 H \left(2 r^2 (H-G)+3\right)}
\end{equation}
\begin{equation}\notag
	\mathcal{A}^{11}=\frac{1}{4 H r^2 (G-H)+4 G-2 H}
\end{equation}
\begin{equation}\notag
	\mathcal{A}^{22}=\frac{1}{e^{-2 G r^2} \left(2 r^2 (G-H)-1\right)+1}
\end{equation}
\begin{equation}\notag
	\mathcal{A}=\frac{2 r^2 e^{2 G r^2}}{2 r^2 (G-H)+e^{2 G r^2}-1}+\frac{e^{2 G r^2}}{4 H r^2 (G-H)+4 G-2 H}-\frac{e^{2 G r^2}}{4 H r^2 (H-G)+6 H}
\end{equation}

\begin{equation}\notag
\mathcal{A}^{0}_{\mkern8mu 0}\mathcal{A}^{00}=-\frac{e^{2 G r^2-2 \left(r^2 (H-G)+P\right)}}{2 H \left(2 r^2 (H-G)+3\right) \left(4 H r^2 (H-G)+6 H\right)}
\end{equation}
\begin{equation}\notag
\mathcal{A}^{1}_{\mkern8mu 1}\mathcal{A}^{11}=\frac{e^{2 G r^2}}{\left(4 H r^2 (G-H)+4 G-2 H\right)^2}
\end{equation}
\begin{equation}\notag
\mathcal{A}^{2}_{\mkern8mu 2}\mathcal{A}^{22}=\frac{r^2 e^{2 G r^2}}{\left(2 r^2 (G-H)+e^{2 G r^2}-1\right) \left(e^{-2 G r^2} \left(2 r^2 (G-H)-1\right)+1\right)}
\end{equation}
\begin{align}\label{rho}
	&\rho=-4 B+\frac{1}{16 \pi }\Bigg[4 e^{-2 G r^2} H \left(-2 G r^2+2 H r^2+3\right)+\frac{2}{r^2}{ e^{-2 G r^2} \left(4 (G-H) H r^4+2 (2 G-3 H) r^2+e^{2 G r^2}-1\right)}+e^{2 G r^2} \times \notag \\
	& \qquad  \left(\frac{2 r^2}{2 (G-H) r^2+e^{2 G r^2}-1}+\frac{1}{4 (G-H) H r^2+4 G-2 H}-\frac{1}{4 H (H-G) r^2+6 H}\right) \alpha +\alpha  \Bigg\{-4 e^{2 \left((H-G) r^2+P\right)} H \times\notag \\
	& \qquad \left(-2 G r^2+4 H r^2-1\right) (\mathcal{A}^{0}_{\mkern8mu 0}\mathcal{A}^{00})+4 \left(2 G^2 r^2+G+\frac{1}{r^2}\right) (\mathcal{A}^{1}_{\mkern8mu 1}\mathcal{A}^{11})+\frac{2}{r}{ e^{2 \left((H-G) r^2+P\right)} \left(G r^2-4 H r^2-1\right) (\mathcal{A}^{0}_{\mkern8mu 0}\mathcal{A}^{00})'}\notag \\
	& \qquad +6 G r (\mathcal{A}^{1}_{\mkern8mu 1}\mathcal{A}^{11})'+2 e^{-2 G r^2} r (\mathcal{A}^{2}_{\mkern8mu 2}\mathcal{A}^{22})'-e^{2 \left((H-G) r^2+P\right)} (\mathcal{A}^{0}_{\mkern8mu 0}\mathcal{A}^{00})''+\Bigg\}-\frac{e^{2 G r^2} \alpha }{H \left(-2 G r^2+2 H r^2+3\right)}\Bigg]-\frac{3}{1-3 \text{$\beta_{1} $}}\times\notag \\
	& \qquad  \Bigg[\beta +4 B \text{$\beta $1}-\frac{1}{16 \pi } (\frac{2 e^{2 G r^2} \alpha  r^2}{2 (G-H) r^2+e^{2 G r^2}-1}-8 e^{2 \left((H-G) r^2+P\right)} H^2 \alpha  (\mathcal{A}^{0}_{\mkern8mu 0}\mathcal{A}^{00}) r^2-2 e^{2 \left((H-G) r^2+P\right)} H \alpha  (\mathcal{A}^{0}_{\mkern8mu 0}\mathcal{A}^{00})' r+\notag \\
	& \qquad 2 H \alpha  (\mathcal{A}^{1}_{\mkern8mu 1}\mathcal{A}^{11})' r +2 e^{-2 G r^2} \alpha  (\mathcal{A}^{2}_{\mkern8mu 2}\mathcal{A}^{22})' r-8 e^{-2 G r^2} H+\frac{3 e^{2 G r^2} \alpha }{4 (G-H) H r^2+4 G-2 H}+8 G \left(H r^2+1\right) \alpha  (\mathcal{A}^{1}_{\mkern8mu 1}\mathcal{A}^{11})+\notag \\
	& \qquad 4 e^{-2 G r^2} \alpha  (\mathcal{A}^{2}_{\mkern8mu 2}\mathcal{A}^{22}) -\frac{2 e^{-2 G r^2}}{r^2}+\frac{2}{r^2}-\frac{e^{2 G r^2} \alpha }{4 H (H-G) r^2+6 H}+\frac{2 \alpha  (\mathcal{A}^{1}_{\mkern8mu 1}\mathcal{A}^{11})'}{r})-\frac{\text{$\beta_{1} $}}{16 \pi } \Bigg(4 e^{-2 G r^2} H \left(-2 G r^2+2 H r^2+3\right)\notag \\
	& \qquad +\frac{2}{r^2}{ e^{-2 G r^2} \left(4 (G-H) H r^4+2 (2 G-3 H) r^2+e^{2 G r^2}-1\right)}{r^2}+e^{2 G r^2} \Bigg(\frac{2 r^2}{2 (G-H) r^2+e^{2 G r^2}-1}\notag \\
	& \qquad +\frac{1}{4 (G-H) H r^2+4 G-2 H}-\frac{1}{4 H (H-G) r^2+6 H}\Bigg) \alpha +\alpha  \Big(-4 e^{2 \left((H-G) r^2+P\right)} H \left(-2 G r^2+4 H r^2-1\right) (\mathcal{A}^{0}_{\mkern8mu 0}\mathcal{A}^{00})\notag \\
	& \qquad +4 \left(2 G^2 r^2+G+\frac{1}{r^2}\right) (\mathcal{A}^{1}_{\mkern8mu 1}\mathcal{A}^{11})+\frac{2}{r}{ e^{2 \left((H-G) r^2+P\right)} \left(G r^2-4 H r^2-1\right) (\mathcal{A}^{0}_{\mkern8mu 0}\mathcal{A}^{00})'}{r}+6 G r (\mathcal{A}^{1}_{\mkern8mu 1}\mathcal{A}^{11})'+2 e^{-2 G r^2} r\times\notag \\
	&\qquad  (\mathcal{A}^{2}_{\mkern8mu 2}\mathcal{A}^{22})' -e^{2 \left((H-G) r^2+P\right)} (\mathcal{A}^{0}_{\mkern8mu 0}\mathcal{A}^{00})''+(\mathcal{A}^{1}_{\mkern8mu 1}\mathcal{A}^{11})''\Big)-\frac{e^{2 G r^2} \alpha }{H \left(-2 G r^2+2 H r^2+3\right)}\Bigg)\Bigg],
\end{align}

\begin{align}\label{pr}
	&p_{r}=-\frac{1}{16 \pi }\Bigg[\frac{2 e^{2 G r^2} \alpha  r^2}{2 (G-H) r^2+e^{2 G r^2}-1}-8 e^{2 \left((H-G) r^2+P\right)} H^2 \alpha  (\mathcal{A}^{0}_{\mkern8mu 0}\mathcal{A}^{00}) r^2-2 e^{2 \left((H-G) r^2+P\right)} H \alpha  (\mathcal{A}^{0}_{\mkern8mu 0}\mathcal{A}^{00})' r+2 H \alpha  (\mathcal{A}^{1}_{\mkern8mu 1}\mathcal{A}^{11})' r\notag \\
	& \qquad +2 e^{-2 G r^2} \alpha  (\mathcal{A}^{2}_{\mkern8mu 2}\mathcal{A}^{22})' r-8 e^{-2 G r^2} H+\frac{3 e^{2 G r^2} \alpha }{4 (G-H) H r^2+4 G-2 H}+8 G \left(H r^2+1\right) \alpha  (\mathcal{A}^{1}_{\mkern8mu 1}\mathcal{A}^{11})+4 e^{-2 G r^2} \alpha  (\mathcal{A}^{2}_{\mkern8mu 2}\mathcal{A}^{22})-\notag \\
	& \qquad \frac{2 e^{-2 G r^2}}{r^2}+\frac{2}{r^2}-\frac{e^{2 G r^2} \alpha }{4 H (H-G) r^2+6 H}+\frac{2 \alpha  (\mathcal{A}^{1}_{\mkern8mu 1}\mathcal{A}^{11})'}{r}\Bigg]-\frac{1}{1-3 \text{$\beta $1}}\Bigg[\beta +4 B \text{$\beta $1}-\frac{1}{16 \pi }\Bigg\{\frac{2 e^{2 G r^2} \alpha  r^2}{2 (G-H) r^2+e^{2 G r^2}-1}-\notag \\
	& \qquad 8 e^{2 \left((H-G) r^2+P\right)} H^2 \alpha  (\mathcal{A}^{0}_{\mkern8mu 0}\mathcal{A}^{00}) r^2-2 e^{2 \left((H-G) r^2+P\right)} H \alpha  (\mathcal{A}^{0}_{\mkern8mu 0}\mathcal{A}^{00})' r+2 H \alpha  (\mathcal{A}^{1}_{\mkern8mu 1}\mathcal{A}^{11})' r+2 e^{-2 G r^2} \alpha  (\mathcal{A}^{2}_{\mkern8mu 2}\mathcal{A}^{22})' r-8 e^{-2 G r^2} H\notag \\
	& \qquad +\frac{3 e^{2 G r^2} \alpha }{4 (G-H) H r^2+4 G-2 H}+8 G \left(H r^2+1\right) \alpha  (\mathcal{A}^{1}_{\mkern8mu 1}\mathcal{A}^{11})+4 e^{-2 G r^2} \alpha  (\mathcal{A}^{2}_{\mkern8mu 2}\mathcal{A}^{22})-\frac{2 e^{-2 G r^2}}{r^2}+\frac{2}{r^2}-\frac{e^{2 G r^2} \alpha }{4 H (H-G) r^2+6 H}\notag \\
	& \qquad +\frac{2 \alpha  (\mathcal{A}^{1}_{\mkern8mu 1}\mathcal{A}^{11})'}{r}\Bigg\}-\frac{1}{16 \pi }\text{$\beta $1} \Bigg[4 e^{-2 G r^2} H \left(-2 G r^2+2 H r^2+3\right)+\frac{2 e^{-2 G r^2} (4 (G-H) H r^4+2 (2 G-3 H) r^2+e^{2 G r^2}-1)}{r^2}\notag \\
	& \qquad +e^{2 G r^2} \left(\frac{2 r^2}{2 (G-H) r^2+e^{2 G r^2}-1}+\frac{1}{4 (G-H) H r^2+4 G-2 H}-\frac{1}{4 H (H-G) r^2+6 H}\right) \alpha +\alpha  \Bigg(-4 e^{2 \left((H-G) r^2+P\right)} H\times \notag \\
	& \qquad  \left(-2 G r^2+4 H r^2-1\right) (\mathcal{A}^{0}_{\mkern8mu 0}\mathcal{A}^{00})+4 \left(2 G^2 r^2+G+\frac{1}{r^2}\right) (\mathcal{A}^{1}_{\mkern8mu 1}\mathcal{A}^{11})+\frac{2 e^{2 \left((H-G) r^2+P\right)} \left(G r^2-4 H r^2-1\right) (\mathcal{A}^{0}_{\mkern8mu 0}\mathcal{A}^{00})'}{r}+\notag \\
	& \qquad 6 G r (\mathcal{A}^{1}_{\mkern8mu 1}\mathcal{A}^{11})'+2 e^{-2 G r^2} r (\mathcal{A}^{2}_{\mkern8mu 2}\mathcal{A}^{22})'-e^{2 \left((H-G) r^2+P\right)} (\mathcal{A}^{0}_{\mkern8mu 0}\mathcal{A}^{00})''+(\mathcal{A}^{1}_{\mkern8mu 1}\mathcal{A}^{11})''\Bigg)-\frac{e^{2 G r^2} \alpha }{H \left(-2 G r^2+2 H r^2+3\right)}\Bigg]\Bigg],
\end{align}

\begin{align}\label{pt}
	&p_{t}=-\frac{1}{1-3 \text{$\beta_{1} $}}\Bigg[\beta +4 B \text{$\beta_{1} $}-\frac{1}{16 \pi }\Bigg\{\frac{2 e^{2 G r^2} \alpha  r^2}{2 (G-H) r^2+e^{2 G r^2}-1}-8 e^{2 \left((H-G) r^2+P\right)} H^2 \alpha  (\mathcal{A}^{0}_{\mkern8mu 0}\mathcal{A}^{00}) r^2-2 e^{2 \left((H-G) r^2+P\right)} H \alpha  (\mathcal{A}^{0}_{\mkern8mu 0}\mathcal{A}^{00})' r\notag \\
	& \qquad +2 H \alpha  (\mathcal{A}^{1}_{\mkern8mu 1}\mathcal{A}^{11})' r+2 e^{-2 G r^2} \alpha  (\mathcal{A}^{2}_{\mkern8mu 2}\mathcal{A}^{22})' r-8 e^{-2 G r^2} H+\frac{3 e^{2 G r^2} \alpha }{4 (G-H) H r^2+4 G-2 H}+8 G \left(H r^2+1\right) \alpha  (\mathcal{A}^{1}_{\mkern8mu 1}\mathcal{A}^{11})+\notag \\
	& \qquad 4 e^{-2 G r^2} \alpha  (\mathcal{A}^{2}_{\mkern8mu 2}\mathcal{A}^{22})-\frac{2 e^{-2 G r^2}}{r^2}+\frac{2}{r^2}-\frac{e^{2 G r^2} \alpha }{4 H (H-G) r^2+6 H}+\frac{2 \alpha  (\mathcal{A}^{1}_{\mkern8mu 1}\mathcal{A}^{11})'}{r}\Bigg\}-\frac{1}{16 \pi }\text{$\beta_{1}$} \Bigg(4 e^{-2 G r^2} H \left(-2 G r^2+2 H r^2+3\right)\notag \\
	& \qquad +\frac{2}{r^2}{ e^{-2 G r^2} \left(4 (G-H) H r^4+2 (2 G-3 H) r^2+e^{2 G r^2}-1\right)}+e^{2 G r^2} \Big(\frac{2 r^2}{2 (G-H) r^2+e^{2 G r^2}-1}+\frac{1}{4 (G-H) H r^2+4 G-2 H}\notag \\
	& \qquad -\frac{1}{4 H (H-G) r^2+6 H}\Big) \alpha +\alpha  \Big(-4 e^{2 \left((H-G) r^2+P\right)} H \left(-2 G r^2+4 H r^2-1\right) (\mathcal{A}^{0}_{\mkern8mu 0}\mathcal{A}^{00})+4 \left(2 G^2 r^2+G+\frac{1}{r^2}\right) (\mathcal{A}^{1}_{\mkern8mu 1}\mathcal{A}^{11})\notag \\
	& \qquad +\frac{2}{r}{ e^{2 \left((H-G) r^2+P\right)} \left(G r^2-4 H r^2-1\right) (\mathcal{A}^{0}_{\mkern8mu 0}\mathcal{A}^{00})'}+6 G r (\mathcal{A}^{1}_{\mkern8mu 1}\mathcal{A}^{11})'+2 e^{-2 G r^2} r (B22)'-e^{2 \left((H-G) r^2+P\right)} (\mathcal{A}^{0}_{\mkern8mu 0}\mathcal{A}^{00})''+\notag \\
	& \qquad (\mathcal{A}^{1}_{\mkern8mu 1}\mathcal{A}^{11})''\Big) -\frac{e^{2 G r^2} \alpha }{H \left(-2 G r^2+2 H r^2+3\right)}\Bigg)\bigg]-\frac{1}{8 \pi }\Bigg[\left(4 G^2 r^2+4 H^2 r^2+2 G+\frac{1}{r^2}\right) \alpha  (\mathcal{A}^{1}_{\mkern8mu 1}\mathcal{A}^{11})+\frac{1}{2} \Big(-2 e^{-2 G r^2} G \alpha  (\mathcal{A}^{2}_{\mkern8mu 2}\mathcal{A}^{22})' r^3\notag \\
	& \qquad +2 e^{-2 G r^2} H \alpha  (\mathcal{A}^{2}_{\mkern8mu 2}\mathcal{A}^{22})' r^3-8 e^{-2 G r^2} H^2 r^2+8 e^{-2 G r^2} G H r^2+\frac{4 e^{2 G r^2} \alpha  r^2}{2 G r^2-2 H r^2+e^{2 G r^2}-1}+e^{-2 G r^2} \alpha  (\mathcal{A}^{2}_{\mkern8mu 2}\mathcal{A}^{22})'' r^2-\notag \\
	& \qquad 2 e^{2 \left((H-G) r^2+P\right)} H \alpha  (\mathcal{A}^{0}_{\mkern8mu 0}\mathcal{A}^{00})' r+6 G \alpha  (\mathcal{A}^{1}_{\mkern8mu 1}\mathcal{A}^{11})' r+6 e^{-2 G r^2} \alpha  (\mathcal{A}^{2}_{\mkern8mu 2}\mathcal{A}^{22})' r+4 e^{-2 G r^2} G-8 e^{-2 G r^2} H+\notag \\
	& \qquad \frac{e^{2 G r^2} \alpha }{4 G \left(H r^2+1\right)-2 H \left(2 H r^2+1\right)}-4 e^{-2 G r^2} \left(G r^2-H r^2-1\right) \alpha  (\mathcal{A}^{2}_{\mkern8mu 2}\mathcal{A}^{22})+\alpha  (\mathcal{A}^{1}_{\mkern8mu 1}\mathcal{A}^{11})''-\frac{e^{2 G r^2} \alpha }{4 H^2 r^2-4 G H r^2+6 H}\Big)\Bigg].
\end{align}

\begin{align}\label{pq}
	&p_{q}=\frac{1}{{1-3 \text{$\beta_{1} $}}}\Bigg[\beta +4 B \text{$\beta_{1} $}-\frac{1}{{16 \pi }}\bigg\{-8 \alpha  H^2 r^2 (\mathcal{A}^{0}_{\mkern8mu 0}\mathcal{A}^{00}) e^{2 (r^2 (H-G)+P)}+8 \alpha  G (\mathcal{A}^{1}_{\mkern8mu 1}\mathcal{A}^{11}) \left(H r^2+1\right)+4 \alpha  (\mathcal{A}^{2}_{\mkern8mu 2}\mathcal{A}^{22}) e^{-2 G r^2} \notag \\
	& \qquad -2 \alpha  H r (\mathcal{A}^{0}_{\mkern8mu 0}\mathcal{A}^{00})' e^{2 \left(r^2 (H-G)+P\right)}+2 \alpha  H r (\mathcal{A}^{1}_{\mkern8mu 1}\mathcal{A}^{11})'+\frac{2}{r}{ \alpha  (\mathcal{A}^{1}_{\mkern8mu 1}\mathcal{A}^{11})'}+2 \alpha  r (\mathcal{A}^{2}_{\mkern8mu 2}\mathcal{A}^{22})' e^{-2 G r^2}+\frac{2 \alpha  r^2 e^{2 G r^2}}{2 r^2 (G-H)+e^{2 G r^2}-1}\notag \\
	& \qquad +\frac{3 \alpha  e^{2 G r^2}}{4 H r^2 (G-H)+4 G-2 H} -\frac{\alpha  e^{2 G r^2}}{4 H r^2 (H-G)+6 H}-8 H e^{-2 G r^2}-\frac{2}{r^{2}}{ e^{-2 G r^2}}+\frac{2}{r^2}\bigg\}-\frac{1}{16 \pi }\bigg\{\text{$\beta_{1} $}\Bigg (\alpha  (-4 H (\mathcal{A}^{0}_{\mkern8mu 0}\mathcal{A}^{00})\times\notag \\
	& \qquad  \left(-2 G r^2+4 H r^2-1\right) e^{2 \left(r^2 (H-G)+P\right)} +4 (\mathcal{A}^{1}_{\mkern8mu 1}\mathcal{A}^{11}) \left(2 G^2 r^2+G+\frac{1}{r^2}\right)+\frac{2}{r}{ (\mathcal{A}^{0}_{\mkern8mu 0}\mathcal{A}^{00})' \left(G r^2-4 H r^2-1\right) e^{2 \left(r^2 (H-G)+P\right)}} \notag \\
	& \qquad +6 G r (\mathcal{A}^{1}_{\mkern8mu 1}\mathcal{A}^{11})'+2 r (\mathcal{A}^{2}_{\mkern8mu 2}\mathcal{A}^{22})' e^{-2 G r^2}-(\mathcal{A}^{0}_{\mkern8mu 0}\mathcal{A}^{00})'' e^{2 \left(r^2 (H-G)+P\right)}+(\mathcal{A}^{1}_{\mkern8mu 1}\mathcal{A}^{11})'')+\alpha  e^{2 G r^2} \bigg(\frac{2 r^2}{2 r^2 (G-H)+e^{2 G r^2}-1}\notag \\
	& \qquad +\frac{1}{4 H r^2 (G-H)+4 G-2 H}-\frac{1}{4 H r^2 (H-G)+6 H}\bigg) -\frac{\alpha  e^{2 G r^2}}{H \left(-2 G r^2+2 H r^2+3\right)}+4 H e^{-2 G r^2} \left(-2 G r^2+2 H r^2+3\right)\notag \\
	& \qquad +\frac{2}{{r^2}}{ e^{-2 G r^2} \left(4 H r^4 (G-H)+2 r^2 (2 G-3 H)+e^{2 G r^2}-1\right)}\Bigg)\bigg\}\Bigg],
\end{align}

\begin{align}\label{rhoq}
	&\rho_{q}=\frac{1}{1-3 \text{$\beta $1}}\Bigg[3 \Bigg\{\beta +4 B \text{$\beta $1}-\frac{1}{16 \pi }\Bigg(-8 \alpha  H^2 r^2 (\mathcal{A}^{0}_{\mkern8mu 0}\mathcal{A}^{00}) e^{2 \left(r^2 (H-G)+P\right)}+8 \alpha  G (\mathcal{A}^{1}_{\mkern8mu 1}\mathcal{A}^{11}) \left(H r^2+1\right)+4 \alpha  (\mathcal{A}^{2}_{\mkern8mu 2}\mathcal{A}^{22}) e^{-2 G r^2}-\notag \\
	& \qquad 2 \alpha  H r (\mathcal{A}^{0}_{\mkern8mu 0}\mathcal{A}^{00})' e^{2 \left(r^2 (H-G)+P\right)}+2 \alpha  H r (\mathcal{A}^{1}_{\mkern8mu 1}\mathcal{A}^{11})'+\frac{2 \alpha  (\mathcal{A}^{1}_{\mkern8mu 1}\mathcal{A}^{11})'}{r}+2 \alpha  r (\mathcal{A}^{2}_{\mkern8mu 2}\mathcal{A}^{22})' e^{-2 G r^2}+\frac{2 \alpha  r^2 e^{2 G r^2}}{2 r^2 (G-H)+e^{2 G r^2}-1}\notag \\
	& \qquad +\frac{3 \alpha  e^{2 G r^2}}{4 H r^2 (G-H)+4 G-2 H}-\frac{\alpha  e^{2 G r^2}}{4 H r^2 (H-G)+6 H}-8 H e^{-2 G r^2}-\frac{2 e^{-2 G r^2}}{r^2}+\frac{2}{r^2}\Bigg)-\frac{1}{16 \pi }\text{$\beta $1} \Bigg(\alpha  \Bigg(-4 H (\mathcal{A}^{0}_{\mkern8mu 0}\mathcal{A}^{00})\notag \\
	& \qquad \left(-2 G r^2+4 H r^2-1\right) e^{2 \left(r^2 (H-G)+P\right)}+4 (\mathcal{A}^{1}_{\mkern8mu 1}\mathcal{A}^{11}) \left(2 G^2 r^2+G+\frac{1}{r^2}\right)+\frac{2 (\mathcal{A}^{0}_{\mkern8mu 0}\mathcal{A}^{00})' \left(G r^2-4 H r^2-1\right) e^{2 \left(r^2 (H-G)+P\right)}}{r}\notag \\
	& \qquad +6 G r (\mathcal{A}^{1}_{\mkern8mu 1}\mathcal{A}^{11})'+2 r (\mathcal{A}^{2}_{\mkern8mu 2}\mathcal{A}^{22})' e^{-2 G r^2}-(\mathcal{A}^{0}_{\mkern8mu 0}\mathcal{A}^{00})'' e^{2 \left(r^2 (H-G)+P\right)}+(\mathcal{A}^{1}_{\mkern8mu 1}\mathcal{A}^{11})''\Bigg)+\alpha  e^{2 G r^2} \bigg(\frac{2 r^2}{2 r^2 (G-H)+e^{2 G r^2}-1}\notag \\
	& \qquad+\frac{1}{4 H r^2 (G-H)+4 G-2 H}-\frac{1}{4 H r^2 (H-G)+6 H}\bigg)-\frac{\alpha  e^{2 G r^2}}{H \left(-2 G r^2+2 H r^2+3\right)}+4 H e^{-2 G r^2} \left(-2 G r^2+2 H r^2+3\right)+\notag \\
	&\qquad \frac{2 e^{-2 G r^2} \left(4 H r^4 (G-H)+2 r^2 (2 G-3 H)+e^{2 G r^2}-1\right)}{r^2}\Bigg)\Bigg\}\Bigg]+4 B.
\end{align}
\begin{align}\label{beta}
	&\beta=(3 \text{$\beta_{1} $}-1) \Bigg[\frac{4 B \text{$\beta_{1} $}}{1-3 \text{$\beta_{1} $}}+\frac{1}{16 \pi }\Bigg(\frac{2 e^{2 G R^2} \alpha  R^2}{2 (G-H) R^2+e^{2 G R^2}-1}-8 e^{2 \left((H-G) R^2+P\right)} H^2 \alpha  (\mathcal{A}^{0}_{\mkern8mu 0}\mathcal{A}^{00}) R^2-2 e^{2 \left((H-G) R^2+P\right)} H \alpha  (\mathcal{A}^{0}_{\mkern8mu 0}\mathcal{A}^{00})' R\notag \\
	& \qquad +2 H \alpha  (\mathcal{A}^{1}_{\mkern8mu 1}\mathcal{A}^{11})' R+2 e^{-2 G R^2} \alpha  (\mathcal{A}^{2}_{\mkern8mu 2}\mathcal{A}^{22})' R-8 e^{-2 G R^2} H+\frac{3 e^{2 G R^2} \alpha }{4 (G-H) H R^2+4 G-2 H}+8 G \left(H R^2+1\right) \alpha  (\mathcal{A}^{1}_{\mkern8mu 1}\mathcal{A}^{11})+\notag \\
	& \qquad 4 e^{-2 G R^2} \alpha  (\mathcal{A}^{2}_{\mkern8mu 2}\mathcal{A}^{22})-\frac{2 e^{-2 G r^2}}{R^2}+\frac{2}{R^2}-\frac{e^{2 G R^2} \alpha }{4 H (H-G) R^2+6 H}+\frac{2 \alpha  (\mathcal{A}^{1}_{\mkern8mu 1}\mathcal{A}^{11})'}{R}\Bigg)-\frac{1}{16 \pi  (1-3 \text{$\beta_{1} $})}\Bigg\{\frac{2 e^{2 G R^2} \alpha  R^2}{2 (G-H) R^2+e^{2 G R^2}-1}\notag \\
	& \qquad -8 e^{2 \left((H-G) R^2+P\right)} H^2 \alpha  (\mathcal{A}^{0}_{\mkern8mu 0}\mathcal{A}^{00}) R^2-2 e^{2 \left((H-G) R^2+P\right)} H \alpha  (\mathcal{A}^{0}_{\mkern8mu 0}\mathcal{A}^{00})' R+2 H \alpha  (\mathcal{A}^{1}_{\mkern8mu 1}\mathcal{A}^{11})' R+2 e^{-2 G R^2} \alpha  (\mathcal{A}^{2}_{\mkern8mu 2}\mathcal{A}^{22})' R-\notag \\
	& \qquad 8 e^{-2 G R^2} H+\frac{3 e^{2 G R^2} \alpha }{4 (G-H) H R^2+4 G-2 H}+8 G \left(H R^2+1\right) \alpha  (\mathcal{A}^{1}_{\mkern8mu 1}\mathcal{A}^{11})+4 e^{-2 G R^2} \alpha  (\mathcal{A}^{2}_{\mkern8mu 2}\mathcal{A}^{22})-\frac{2 e^{-2 G R^2}}{R^2}+\frac{2}{R^2}-\notag \\
	& \qquad \frac{e^{2 G R^2} \alpha }{4 H (H-G) R^2+6 H}+\frac{2 \alpha  (\mathcal{A}^{1}_{\mkern8mu 1}\mathcal{A}^{11})'}{R}\Bigg\}-\frac{\beta_{1}}{16 \pi  (1-3 \text{$\beta_{1}$})} \Bigg(4 e^{-2 G R^2} H \left(-2 G R^2+2 H R^2+3\right)+\notag \\
	& \qquad \frac{2}{R^2}{ e^{-2 G R^2} \left(4 (G-H) H R^4+2 (2 G-3 H) R^2+e^{2 G R^2}-1\right)}+e^{2 G R^2} \bigg(\frac{2 R^2}{2 (G-H) r^2+e^{2 G R^2}-1}+\frac{1}{4 (G-H) H R^2+4 G-2 H}\notag \\
	& \qquad -\frac{1}{4 H (H-G) R^2+6 H}\bigg) \alpha +\alpha  \Bigg(-4 e^{2 \left((H-G) R^2+P\right)} H \left(-2 G R^2+4 H R^2-1\right) (\mathcal{A}^{0}_{\mkern8mu 0}\mathcal{A}^{00})+4 \left(2 G^2 R^2+G+\frac{1}{R^2}\right) (\mathcal{A}^{1}_{\mkern8mu 1}\mathcal{A}^{11})\notag \\
	& \qquad +\frac{2}{R}{ e^{2 \left((H-G) R^2+P\right)} \left(G R^2-4 H R^2-1\right) (\mathcal{A}^{0}_{\mkern8mu 0}\mathcal{A}^{00})'}+6 G R (\mathcal{A}^{1}_{\mkern8mu 1}\mathcal{A}^{11})'+2 e^{-2 G rR2} R (\mathcal{A}^{2}_{\mkern8mu 2}\mathcal{A}^{22})'-e^{2 \left((H-G) r^2+P\right)} (\mathcal{A}^{0}_{\mkern8mu 0}\mathcal{A}^{00})''\notag \\
	& \qquad +(\mathcal{A}^{1}_{\mkern8mu 1}\mathcal{A}^{11})''\Bigg)-\frac{e^{2 G R^2} \alpha }{H \left(-2 G R^2+2 H R^2+3\right)}\Bigg)\Bigg]
\end{align}

\section*{Data Availability Statement}
\hskip\parindent
\small
The research presented in this article is solely based on pure mathematics, and no external data was used.

\section*{Conflict of Interest}
\hskip\parindent
\small
The authors assert that they do not have any conflicts of interest.

\section*{Contributions}
\hskip\parindent
\small
We hereby affirm that each author has contributed equally to this manuscript.

\section*{Data Availability Statement}
The authors declare that the data supporting the findings of this study are available within the article.

\end{document}